\documentclass[pra,twocolumn,superscriptaddress]{revtex4}

\usepackage{amsmath}
\usepackage{latexsym}
\usepackage{amssymb}
\usepackage{pdfpages}
\usepackage{graphicx}
\usepackage{subfigure}
\usepackage[colorlinks=true, citecolor=blue, urlcolor=blue]{hyperref}
\usepackage{float}
\usepackage{epstopdf}
\usepackage{pgffor}
\usepackage{amsfonts}
\usepackage{textcomp}
\usepackage{comment}
\usepackage{braket}
\usepackage{bbm}
\usepackage{xcolor}
\definecolor{myurlcolor}{rgb}{0,0,0.8}
\definecolor{mycitecolor}{rgb}{0,0,0.8}
\definecolor{myrefcolor}{rgb}{0,0,0.8}
\usepackage{hyperref}
\usepackage[capitalise]{cleveref}
\hypersetup{colorlinks,
linkcolor=myrefcolor,
citecolor=mycitecolor,
urlcolor=myurlcolor}

\usepackage[draft]{fixme}
\usepackage{amsmath,bbm}
\usepackage{graphicx}
\usepackage{amsfonts}
\usepackage{amssymb}
\usepackage{amsmath, amssymb, amsthm,verbatim,graphicx,bbm}
\usepackage{mathrsfs}
\usepackage{color,xcolor,longtable}

\makeatletter
\usepackage{pifont}
\makeatother

\usepackage{epstopdf}
\usepackage{subfigure}
\usepackage{appendix}

\makeatletter
\usepackage{pifont}
\makeatother

\usepackage{dcolumn}
\usepackage{epstopdf}

\begin{document}

\title{Quantum parameter estimation of non-Hermitian systems with optimal measurements}
\author{Xinglei Yu}
\author{Chengjie Zhang}
\email{chengjie.zhang@gmail.com}
\affiliation{School of Physical Science and Technology, Ningbo University, Ningbo, 315211, China}

\begin{abstract}
  Quantum parameter estimation with Hermitian systems has been applied in various fields, but there are relatively few results concerning non-Hermitian systems. Here, we study the quantum parameter estimation for general non-Hermitian Hamiltonians and derive an intuitive expression of quantum Fisher information (QFI) for pure states. Furthermore, we propose the condition for optimal measurements, which is applicable to both Hermitian and non-Hermitian Hamiltonians. To illustrate these results, we calculate and study the QFI of a specific $\mathcal{PT}$-symmetric non-Hermitian Hamiltonian, and give the optimal measurement. Surprisingly, we find some interesting properties of this $\mathcal{PT}$-symmetric Hamiltonian QFI, such as the mutations in QFI at EP. Moreover, we also compare the variance of estimation generated by the optimal measurement with the theoretical precision bound to verify the condition for optimal measurements we proposed.
\end{abstract}
\date{\today}

\maketitle

\section{Introduction}
Quantum parameter estimation is a theory committed to high-precision measurement of parameters in a quantum system, which plays a crucial role in modern science and engineering. Cram\'{e}r-Rao bound (CRB) and Fisher information are of great importance in classical parameter estimation, which characterize and limit the estimation precision. Helstrom \cite{rev1} and Holevo \cite{rev2} proposed the parameter-based uncertainty relation, and pioneered quantum metrology based on quantum parameter estimation. After that, Braunstein \emph{et al}. \cite{rev3,rev4} extended Fisher information to the quantum regime, and proposed quantum Fisher information (QFI) which is the optimal Fisher information over different valid quantum measurements. It is a metric that characterizes the maximum amount of attainable information about the given parameter and provides a new ultimate bound often called quantum CRB.

With the development of technology, precision measurement has attracted extensive attention in various fields, which has also promoted the development of quantum parameter estimation theory. Recently, scientists have found that QFI played an essential role in quantum information theory, especially in quantum entanglement theory. The QFI can be used as criteria for bipartite entanglement \cite{rev5} and multipartite entanglement, it has been proved that only genuine multipartite entanglement is able to reach a maximal metrological sensitivity \cite{rev6,rev7}. The QFI was also discovered to be related to the speed of quantum evolution \cite{rev8,rev9,rev10} and used for the proper definition of a macroscopic superposition \cite{rev11,rev12}. A large multipartite entanglement with large QFI can evolve faster than a state with a small QFI. In addition, QFI has also been applied to quantum Zeno effect \cite{rev13,rev14}.

In recent years, quite a few studies have shown that non-Hermitian systems have many interesting properties which attracts considerable attention. An important discovery is that a non-Hermitian Hamiltonian with exact parity-time (PT) symmetry may exhibit entirely real spectrum \cite{rev15,rev16,rev17,rev18,rev19}. In such $\mathcal{PT}$-symmetric systems, eigenstates are not orthogonal, and states do not conserve probabilities after evolution. Interestingly, the $\mathcal{PT}$-symmetry may be suddenly broken once the non-Hermiticity parameter exceeds a certain critical value \cite{rev20,rev21}. These discoveries have promoted the new development in material science \cite{rev22,rev23}, topology \cite{rev24,rev25} and especially optics \cite{rev26,rev27,rev28,rev29,rev30}.

In previous quantum metrology researches, there have been a great quantity of works on Hermitian systems \cite{hev1,hev2,hev3,hev4,hev5,hev6,hev7,hev8,hev9,hev10,hev11,hev12,hev13,hev14}. However, only a few quantum sensors based on non-Hermitian systems have been proposed recently \cite{rev31,rev32,rev33,rev34,nev1,nev2,nev3,nev4,nev5,nev6,nev7}, these researches show that sensors may have enhanced sensitivity in $\mathcal{PT}$-symmetric non-Hermitian systems. The optimal homodyne-based measurement scheme and fundamental bounds of non-Hermitian quantum sensor has been studied in Ref. \cite{opm1}, it considers the effect of noise as induced by non-Hermitianity. And the parameter estimation in $\mathcal{PT}$ symmetrical cavity magnonics system has also been studied \cite{emp1}. Recently, the quantum Cramer-Rao bound has been also extend to the non-Hermitian regime with non-Hermitian Symmetric Logarithmic Derivative \cite{emnh}. Considering the new developments in physics arising from the studies of non-Hermitian systems over the past two decades, it is significant to promote the study of quantum metrology in non-Hermitian systems.

In this paper, we derive a general intuitive expression of QFI for arbitrary parameter-independent non-Hermitian Hamiltonians and find the condition for optimal measurements based on the error-propagation formula. According to the expression we proposed, we calculate and study the QFI of a specific $\mathcal{PT}$-symmetric non-Hermitian Hamiltonian and find the optimal measurement. Interestingly, we find that the QFI of this $\mathcal{PT}$-symmetric Hamiltonian is discontinuous at EP, it will reduce to zero as EP is approached, but changes suddenly when the EP is exactly reached. We also find that the channel QFI \cite{cqfi} which is maximized after performing the input optimization does not tend to zero near EP, and the expressions of the channel QFI are the same in both cases of $\mathcal{PT}$-symmetry broken and not. Comparing the estimation variance generated by the optimal measurement with the lower bound of quantum CRB, we verify the condition for optimal measurement that we proposed.

\section{QFI for general non-Hermitian Hamiltonians}
Before we calculate the QFI for parameter-independent non-Hermitian Hamiltonians, let us first review some basic concepts of the quantum estimation theory. The precision of estimation is usually characterized by the variance of the estimator, so we expect to reduce the variance as much as possible. But how can we know the lower bound of variance? According to the estimation theory, the variance of any unbiased estimator $\hat{\theta}$ is limited by quantum CRB $V(\hat{\theta})\geq1/(n\mathcal{F}_\theta)$, where $n$ is the number of measurements and $\theta$ is the actual parameter which we expect to estimate, $\hat{\theta}$ is the estimator of $\theta$, and $\mathcal{F}_\theta$ is the quantum Fisher information which gives the theoretical ultimate bound of parameter estimation precision. Note that the quantum CRB is asymptotically achievable, which requires the number of measurements $n$ to be large. The general expression of QFI is $\mathcal{F}_\theta=\mathrm{Tr}[L_{\theta}^{2}\rho_{\theta}]$, where $L_{\theta}$ is a self-adjoint operator called Symmetric Logarithmic Derivative (SLD), satisfies the equation $(L_{\theta}\rho_{\theta}+\rho_{\theta}L_{\theta})/2=\partial_{\theta}\rho_{\theta}$. For an arbitrary state written in its eigenbasis $\rho_{\theta}=\sum_{n}c_n\ket{\psi_n}\bra{\psi_n}$, the expression of QFI is \cite{rev1,rev2}
\begin{eqnarray}\label{a1}
\mathcal{F}_\theta=2\sum_{n,m}\frac{|\bra{\psi_n}\partial_{\theta}\rho_{\theta}\ket{\psi_m}|^2}{c_n+c_m},
\end{eqnarray}
where $c_n+c_m$ is required not to be zero. For unitary families and pure states model, the final state is $\rho_{\theta}=|\psi_\theta\rangle\langle\psi_\theta|=\hat{U}_{\theta}\rho_{0}\hat{U}_{\theta}^{\dagger}$ and the QFI is $\mathcal{F}_\theta=4\bra{\psi_{0}}(\Delta \hat{H})^{2}\ket{\psi_{0}}$, where $\hat{U}_{\theta}=e^{-i\hat{H}\theta}$ is an unitary operator and $\hat{H}$ is the corresponding Hermitian generator.

Now let us discuss the QFI for parameter-independent non-Hermitian Hamiltonians in the case of pure states. If we turn to non-Hermitian Hamiltonians, the evolution operator is no more unitary due to $\hat{H}\neq \hat{H}^{\dagger}$, and the final state $\ket{\psi_{\theta}}$ is no longer normalized after evolution. Therefore, the QFI for pure states may no longer be written as the variance of Hamiltonian. According to the Born rule, we have $p(x|\theta)=\mathrm{Tr}[\pi_{x} \rho_{\theta}]$, this requires $\rho_{\theta}$ to be normalized, so we cannot directly substitute $\rho_{\theta}$ into Eq. (\ref{a1}) to get the result. As for measurement in non-Hermitian systems, the probabilities of measurement outcomes always sum up to $1$, even if there are gains and losses during evolution process. For instance, the number of photons is reduced from $N_0$ to $N^\prime=N_{0}\langle\psi_\theta|\psi_\theta\rangle$ after non-unitary evolution, the final states are measured with a set of projection operator $\Pi=\sum_i|i\rangle\langle i|$, then the total probability is $\sum_i P_i=\sum_i N_i/N^\prime=(N_0\sum_i\langle\psi_\theta|i\rangle\langle i|\psi_\theta\rangle)/(N_0\langle\psi_\theta|\psi_\theta\rangle)=1$, where $N_i$ represents the number of photons whose measurement outcome is $|i\rangle$. Thus, we assume that the pure state $\tilde{\rho}_{\theta}$ is the normalized final state :
\begin{eqnarray}
\tilde{\rho}_{\theta}=|\varphi_{\theta}\rangle\langle\varphi_{\theta}|=\frac{\rho_{\theta}}{K_{\theta}},
\end{eqnarray}
where $K_{\theta}$ is the normalization coefficient, and $\ket{\varphi_{\theta}}=(\hat{U}_{\theta}/\sqrt{K_{\theta}})\ket{\psi_{0}}$. Note $K_{\theta}$ is the inner product of $\ket{\psi_{\theta}}$ itself, it must be non-negative. According to Eq. (\ref{a1}), for the normalized $\tilde{\rho}_{\theta}$, we derive the QFI as follows,
\begin{eqnarray}
\mathcal{F}_\theta=4(\langle \hat{H}^{\dagger}\hat{H}\rangle_\theta-\langle \hat{H}^{\dagger}\rangle_\theta\langle \hat{H}\rangle_\theta).\label{a3}
\end{eqnarray}
Here, we define $\langle \hat{M}\rangle_\theta$ as the expectation value of the operator $\hat{M}$ on the normalized final state $\ket{\varphi_\theta}$ ($\langle \hat{M}\rangle_\theta=\bra{\varphi_{\theta}}\hat{M}\ket{\varphi_{\theta}}$). The detailed derivation can be found in Appendix A. This expression is valid under the assumption that the evolution of non-Hermitian system satisfies the Schr\"{o}dinger equation: $i\partial_t\ket{\psi}=\hat{H}\ket{\psi}$. For open systems, if the non-Hermitian Hamiltonian can be regard as the effective description of the open system \cite{qtj}, our result is also valid. This result is quite similar to QFI of Hermitian Hamiltonian, which is four times the variance of non-Hermitian generator, but here, the QFI depends on the normalized final state $\tilde{\rho}_{\theta}$. Obviously, the QFI is the parameter-dependent function, we can infer that for given non-Hermitian Hamiltonians and initial states, the QFI is no more a constant, but changes with the parameter $\theta$.

Note that the QFI we proposed above is based on a successful detection event, it does not represent the ultimate estimation precision for given resource of probe states. For example, if we input $N_0$ probe states, the total number of final states that we are able to detect is $N'=K_\theta N_{0}$. According to the quantum Cram\'{e}r-Rao bound, the ultimate estimation precision is $V(\hat{\theta})\geq1/(N'\mathcal{F}_\theta)$. Therefore, for given input resource, we could define $\mathcal{I}_\theta=K_\theta\mathcal{F}_\theta$ to characterize to estimation precision, it indicates that the gain and loss during evolution will effect the estimation precision. As for the specific implementation of non-Hermitian system, the evolution could be a little different from the real theoretical non-Hermitian system evolution, e.g., the effective evolution could be $\hat{U}'_\theta=f(\theta)\hat{U}_\theta$ \cite{spn1,spn2}, where $f(\theta)$ is a function. To characterize the estimation precision of effevtive implementations of non-Hermitian systems, we just need to replace $\hat{U}_\theta$ with $\hat{U}'_\theta$ to calculate the normalization coefficient.

\section{Condition for optimal measurements}
The quantum CRB and QFI give the achievable ultimate bound on precision of parameter estimation. However, how can we actually reach this bound in parameter-independent non-Hermitian systems? It is well-known that the optimal measurement is obtained if we measure in the eigenbasis of SLD, and the concise expression for SLD $L_\theta=2(|\partial_\theta\varphi_\theta\rangle\langle\varphi_\theta|+|\varphi_\theta\rangle\langle\partial_\theta\varphi_\theta|)$ has been proposed already, it is also valid for non-Hermitian Hamiltonians. However, SLD is not unique, i.e., the optimal measurement is not unique, the above expression cannot find all SLD. Here, we propose an alternative condition which also has a wide scope of application, and we prove that $L_\theta=2(|\partial_\theta\varphi_\theta\rangle\langle\varphi_\theta|+|\varphi_\theta\rangle\langle\partial_\theta\varphi_\theta|)$ satisfies the condition and find a new SLD in Appendix B. And the condition for optimal measurements also indicates the connection between optimal measurements and uncertain relation. We first define $\delta \hat{M}=\hat{M}-\langle \hat{M}\rangle_{\theta}$ and $(\Delta\hat{M})^2=\langle\delta\hat{M^\dagger}\delta\hat{M}\rangle_\theta=\langle\hat{M}^{\dagger}\hat{M}\rangle_\theta-\langle\hat{M}^{\dagger}\rangle_\theta\langle\hat{M}\rangle_\theta$. For an observable $\hat{A}$, the precision of estimation can be characterized with the error-propagation formula \cite{ero1,ero2},
\begin{eqnarray}
(\Delta\theta)^2=\frac{(\Delta \hat{A})^2}{n|\partial_\theta \langle \hat{A}\rangle_\theta|^2}. \label{er}
\end{eqnarray}
Similarly, the error-propagation formula is also asymptotically right when $n$ is large. According to the Non-Hermitian uncertainty relationship $(\Delta \hat{A})^2(\Delta \hat{B})^2\geq|\langle \hat{A}^{\dagger}\hat{B}\rangle-\langle \hat{A}^{\dagger}\rangle\langle \hat{B}\rangle|$ \cite{ur1,ur2,ur3,rev35,zxev35} which was first proposed by Pati in Ref. \cite{ur1}, we can obtain an inequality as follows,
\begin{eqnarray}
(\Delta\theta)^2=\frac{(\Delta \hat{A})^2}{n|\partial_\theta \langle \hat{A}\rangle_\theta|^2}\geq\frac{(\Delta \hat{A})^2}{4n(\Delta \hat{A})^2(\Delta \hat{H})^2}=\frac{1}{n\mathcal{F}_{\theta}}, \label{a4}
\end{eqnarray}
the right term is exactly the lower bound of quantum CRB. To saturate this inequality, the Hermitian operator $\hat{A}$ must obey the relationship as follows,
\begin{eqnarray}
\ket{f}=iC\ket{g}, \label{a5}
\end{eqnarray}
where $\ket{f}=\delta \hat{H}\ket{\varphi_\theta}$, $\ket{g}=\delta \hat{A}\ket{\varphi_\theta}$ and $C$ is a real number. Notice that Eq. (\ref{a5}) ensures the saturation of uncertain relationships of $\hat{A}$ and $\hat{H}$, which indicates that the saturation of uncertain relationships is an underlying requirement to achieve quantum CRB. Except for the general Non-Hermitian Hamiltonians, this relationship also applies to the Hermitian Hamiltonians, which has already been obtained in Ref. \cite{rev36}. The detailed derivation of this relationship can be found in the Appendix B. Note that this relationship not only depends on the measurements, but is also related to the initial states. For a given Hamiltonian, on the basis of finding the optimal initial states to maximize the QFI, we can further find the optimal measurements according to Eq. (\ref{a5}). In this way, we can estimate the parameters with highest precision.

\section{Example of a $\mathcal{PT}$-symmetric Hamiltonian}
Let us now consider a two-level $\mathcal{PT}$-symmetric Hamiltonian to illustrate the results in previous sections. We calculate the QFI in both cases of $\mathcal{PT}$-symmetry broken and not, the details can be found in the Appendix C. The $\mathcal{PT}$-symmetric Hamiltonian is \cite{rev17}
\begin{eqnarray}
\hat{H}_{s}=\begin{pmatrix} re^{i\omega} & s \\ s & re^{-i\omega}  \end{pmatrix},\label{b1}
\end{eqnarray}
where $s>0$, $r$ and $\omega$ are real. We can see that the eigenvalues $\lambda_{\pm}=\mu\pm\sqrt{\nu_0}=r\cos{\omega}\pm\sqrt{s^2-r^2\sin^2{\omega}}$ are still real when the $\mathcal{PT}$-symmetry is not broken, i.e., $s^2>r^2\sin^2{\omega}$. Note that at EP, i.e., $s^2=r^2\sin^2{\omega}$, the eigenvalues $\lambda=r\cos\omega$ are degenerate, and when $s^2<r^2\sin^2{\omega}$, $\mathcal{PT}$-symmetry will spontaneously be broken, the two eigenvalues $\varepsilon_{\pm}=\mu\pm i\sqrt{\nu_1}=r\cos{\omega}\pm i\sqrt{r^2\sin^2{\omega}-s^2}$ are complex.

First we calculate the QFI of Hamiltonian $\hat{H}_s$ when $\mathcal{PT}$-symmetry is not broken. Assume that the eigenstates are normalized, we have
\begin{eqnarray}
\ket{\lambda_+}=\frac{1}{\sqrt{2}}\begin{pmatrix}e^{i\alpha/2} \\ e^{-i\alpha/2} \end{pmatrix},\quad
\ket{\lambda_-}=\frac{i}{\sqrt{2}}\begin{pmatrix}e^{-i\alpha/2} \\ -e^{i\alpha/2} \end{pmatrix},\nonumber
\end{eqnarray}
where $\sin{\alpha}=(r/s)\sin{\omega}$, and we set the initial state to be normalized and arbitrary
\begin{eqnarray}
\ket{\psi_{0}}=N(\ket{\lambda_{+}}+me^{i\varphi}\ket{\lambda_{-}}),\nonumber
\end{eqnarray}
where $m$ and relative phase $\varphi$ are real, $N$ is the normalization coefficient. Notice that the eigenstates are not orthogonal, we have $\langle\lambda_{+}|\lambda_{-}\rangle=\sin\alpha$. We apply Eq. (\ref{a3}) and derive the QFI as follows:
\begin{eqnarray}\label{f1}
\mathcal{F}_{\theta}^{(r)}=\frac{16m^2\nu_0^2}{[(1+m^2)s+2mr\sin\omega\cos(2\sqrt{\nu_0}\theta+\varphi)]^2}.
\end{eqnarray}
The expression shows that QFI will oscillate with $\theta$ due to $\cos(2\sqrt{\nu_0}\theta+\varphi)$.

Obviously, the value of QFI would be affected by the initial states. We further find out the optimal initial state that maximizes the QFI, which is also called channel QFI. Notice the form of Eq. (\ref{f1}), we can see that the relative phase $\varphi$ is only contained in $\cos(2\sqrt{\nu_0}\theta+\varphi)$. Thus, $\varphi$ has nothing to do with the amplitude and frequency of $\mathcal{F}_{\theta}^{(r)}$. Then we solve $\partial\mathcal{F}_{\theta}^{(r)}/\partial m=0$, and obtain that the extremum of $\mathcal{F}_{\theta}^{(r)}$ reaches the maximum when $m=\pm1$. Now we can make sure that in the case of PT-symmetry not broken, the optimal initial state is
\begin{eqnarray}
\ket{\psi_{0}}=N(\ket{\lambda_{+}}\pm e^{i\varphi}\ket{\lambda_{-}}).
\end{eqnarray}

\begin{figure}[t]
\centering
\includegraphics[ width=0.5\textwidth]{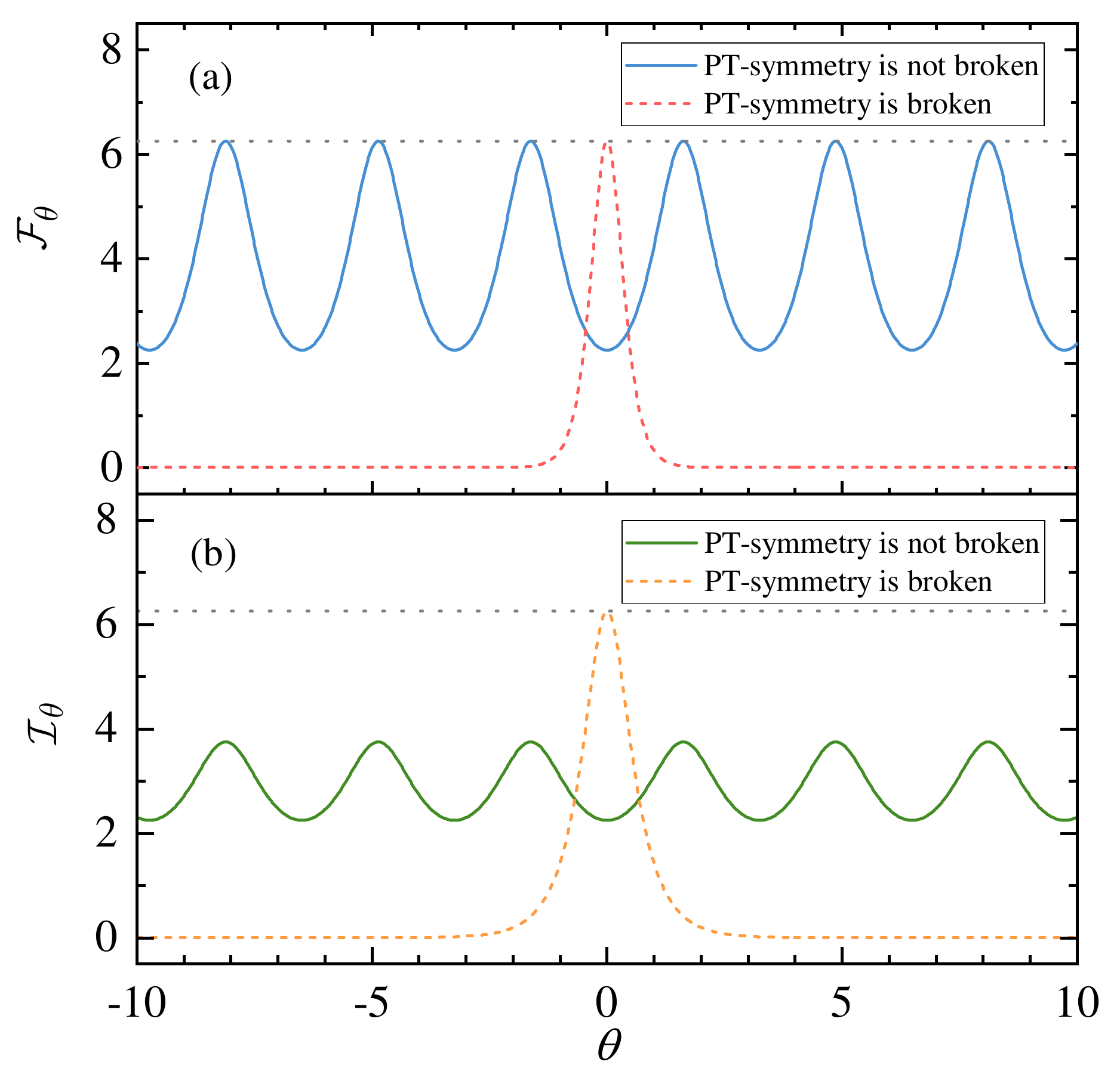}
\caption{The evolution of QFI and $\mathcal{I}_\theta$ as functions of $\theta$. (a) The blue solid curve corresponds to the QFI when $\mathcal{PT}$-symmetry is not broken, we set $s=1,r=0.25,\omega=\pi/2$, and the initial state is $\ket{\psi_0}=N(\ket{\lambda_+}+\ket{\lambda_-})$. The red dashed curve corresponds to the QFI in the case of $\mathcal{PT}$-symmetry broken, we set $s=0.25,r=1,\omega=\pi/2$, and the initial state is $\ket{\psi_0}=N(\ket{\varepsilon_+}-\ket{\varepsilon_-})$. Thus, these two systems have the same modulus of energy, but $\sin\alpha=1/\kappa$. According to Eq. (\ref{max2}), their maximum value of QFI both are 6.25. (b) The green solid curve corresponds to the QFI when $\mathcal{PT}$-symmetry is not broken, and the orange dashed curve corresponds to the QFI in the case of $\mathcal{PT}$-symmetry broken. The parameters and initial states in both cases are the same as (a).} \label{fig1}
\end{figure}

\begin{figure}[t]
\centering
\includegraphics[width=0.5\textwidth]{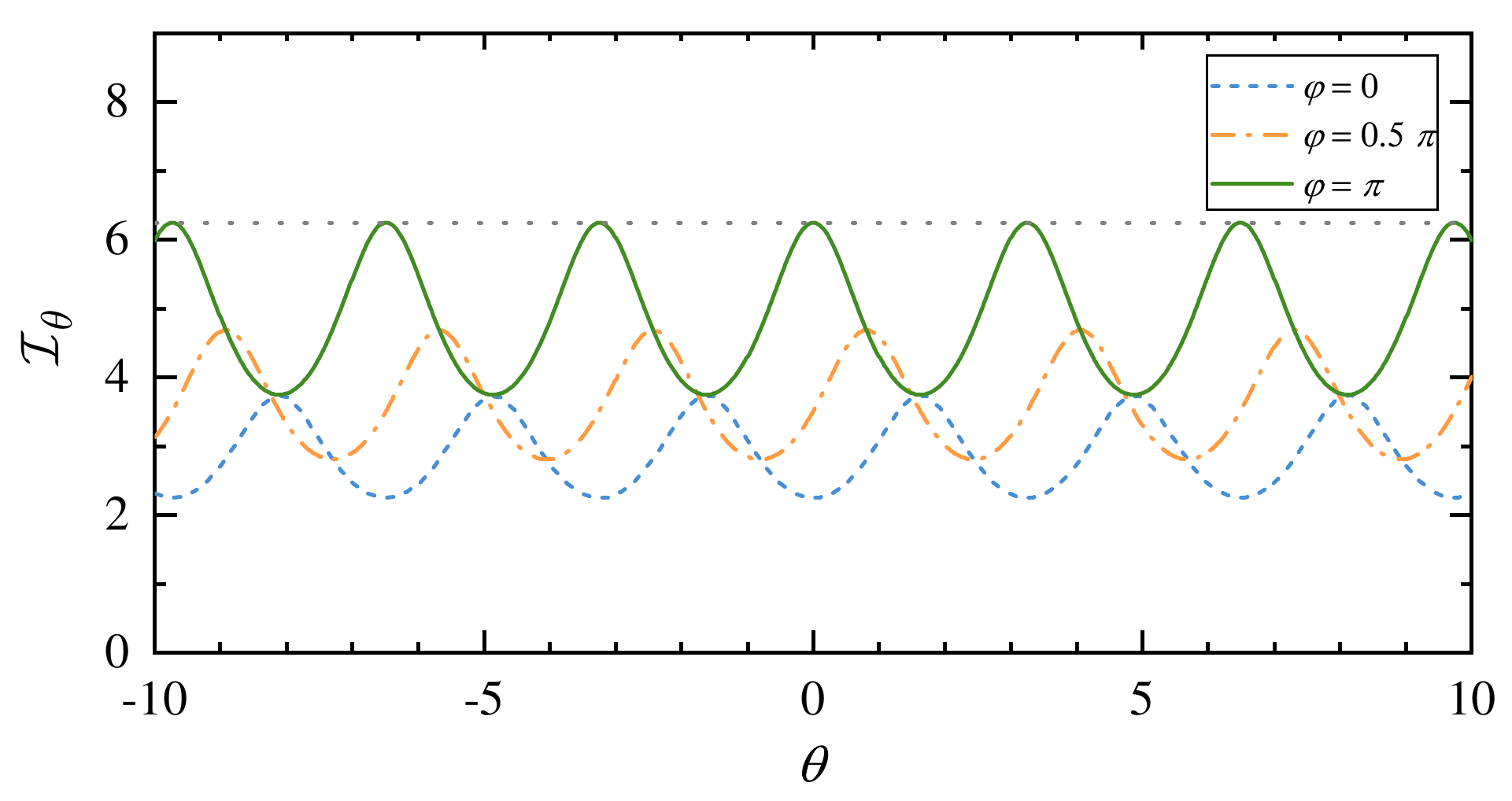}
\caption{The evolution of $\mathcal{I}_{\theta}$ in the case of $\mathcal{PT}$-symmetry not broken as a function of $\theta$. We set that $s=1$, $r=0.4$, $\omega=\pi/2$, $m=1$ and $\varphi$ changes from $0$ to $\pi$, where blue dashed line ($\varphi=0$), orange dot-dashed line ($\varphi=0.5\pi$) and green solid line ($\varphi=\pi$).}\label{IFIphi}
\end{figure}

Next we are going to calculate the QFI of $\hat{H}_s$ when $\mathcal{PT}$-symmetry is broken. In this case, the normalized eigenstates are
\begin{eqnarray}
\ket{\varepsilon_{\pm}}&=&\frac{1}{\sqrt{(r\sin\omega\pm\sqrt{\nu_1})^2+s^2}}\begin{pmatrix} i(r\sin\omega\pm\sqrt{\nu_1}) \\ s \end{pmatrix},\nonumber
\end{eqnarray}
and the arbitrary initial state is $\ket{\psi_{0}}=N(\ket{\varepsilon_{+}}+me^{i\varphi}\ket{\varepsilon_{-}})$. Notice that in this case, $(r/s)\sin{\omega}>1$, we cannot define $\alpha$ as before, here we use $\kappa=(r/s)\sin{\omega}$. According to Eq. (\ref{a3}), the corresponding QFI is
\begin{eqnarray}
\mathcal{F}_{\theta}^{(i)}=\frac{16m^2\nu_1^2e^{4\sqrt{\nu_1}\theta}}{[A(e^{4\sqrt{\nu_1}\theta}+m^2)+2mse^{2\sqrt{\nu_1}\theta}\cos\varphi]^2},\label{f2}
\end{eqnarray}
where $A=|r\sin\omega|$. We also find the optimal initial state by solving $\partial\mathcal{F}_{\theta}^{(i)}/\partial\varphi=0$ and $\partial\mathcal{F}_{\theta}^{(i)}/\partial m=0$. The optimal initial state is
\begin{eqnarray}
\ket{\psi_{0}}=N(\ket{\varepsilon_{+}}+m\ket{\varepsilon_{-}}),
\end{eqnarray}
where $m<0$ and $\varphi=0$. We can see that the Eq. (\ref{f2}) is quite different from the Eq. (\ref{f1}), without the oscillation term but gain and attenuation term. As shown in Fig. {\ref{fig1}} (a), QFI oscillates with $\theta$ in the case of $\mathcal{PT}$-symmetry not broken; when $\mathcal{PT}$-symmetry is broken, QFI has only one peak and quickly attenuates with $\theta$. Therefore, the QFI performance is better when the $\mathcal{PT}$-symmetry is not broken, and we do more research on this. We also find that in the case of $\mathcal{PT}$-symmetry not broken, the curve of QFI shifts along the $\theta$-axis as we change the relative phase $\varphi$ in initial states $\ket{\psi_0}$. So we can maximize the QFI for a specific parameter $\theta_0$ by adjusting the relative phase $\varphi$ in initial states. Moreover, we also plot $\mathcal{I}_\theta$ in both cases of $\mathcal{PT}$-symmetry broken and not broken. As shown in Fig. \ref{fig1} (b), the amplitude of $\mathcal{I}_\theta$ is smaller than the QFI when $\mathcal{PT}$-symmetry is not broken, but the maximum value of $\mathcal{I}_\theta$ does not change when $\mathcal{PT}$-symmetry is broken. That is due to the introduction of $K_\theta$ which make the relative phase $\varphi$ also be able to effect the amplitude, as shown in Fig. \ref{IFIphi}, the maximum value of $I_\theta$ can also reach 6.25 when $\varphi=\pi$.

With the optimal initial states, we calculate the channel QFI in both cases of $\mathcal{PT}$-symmetry broken and not. Interestingly, we find that the channel QFI $\mathcal{F}_{\theta,max}$ has the same expressions in both cases,
\begin{eqnarray}
  \mathcal{F}_{\theta,max}=4(s+|r\sin\omega|)^2, \quad &(s^2\neq r^2\sin^2\omega).\label{max2}
\end{eqnarray}

\begin{figure}[t]
  \centering
  \includegraphics[ width=0.2375\textwidth]{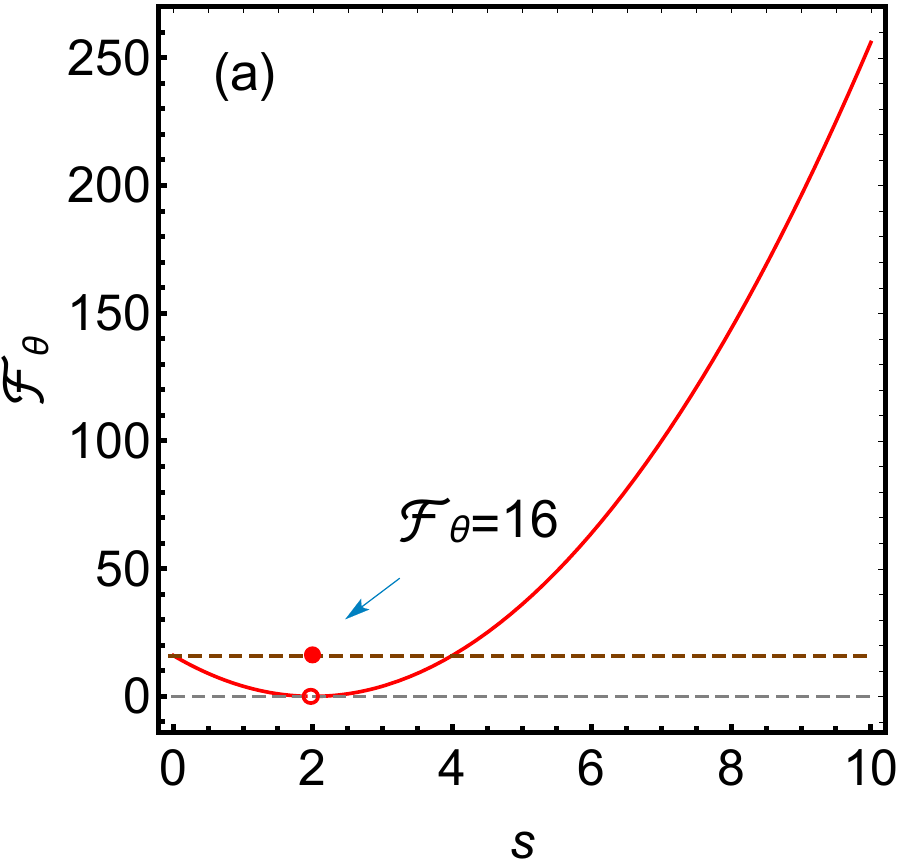}
  \includegraphics[ width=0.2375\textwidth]{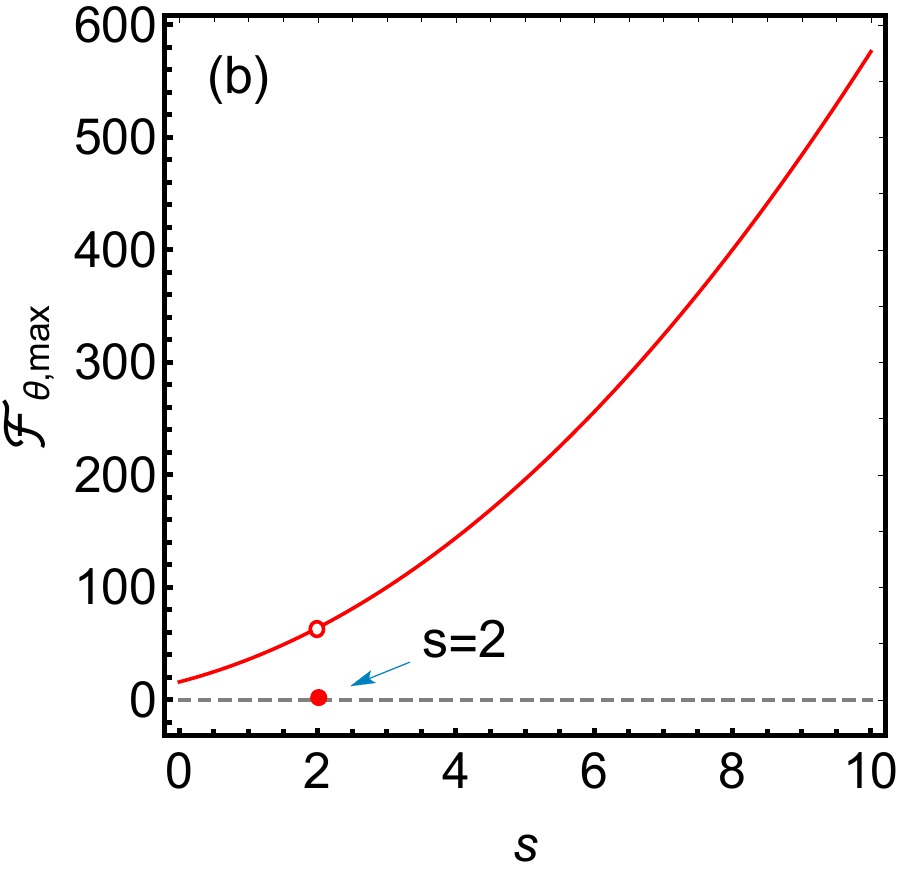}
  \caption{The evolutions of QFI as a function of $s$. (a) We set $r=2,\omega=\pi/2,m=1,\varphi=0$ and $\theta=0$, hence the system reaches EP when $s=2$. And the initial state is $|0\rangle\langle0|$ at EP, we have $\mathcal{F}_{\theta}=16$. (b) We set $r=2$ and $\omega=\pi/2$, the system reaches EP when $s=2$.}\label{fig2}
\end{figure}

As for EP, the eigenvalues are degenerate $\lambda_\pm=\mu$, if the initial states can be expressed with eigenstates, we have
\begin{eqnarray}
\mathcal{F}_{\theta}&=&4(\bra{\psi_{0}}e^{i\mu\theta}\mu^{2}e^{-i\mu\theta}\ket{\psi_{0}}-\bra{\psi_{0}}e^{i\mu\theta}\mu e^{-i\mu\theta}\ket{\psi_{0}}^{2})\nonumber\\
&=&4(\mu^{2}-\mu^{2})=0.\nonumber
\end{eqnarray}
However, the eigenstates also coalesce at EP, they are not a set of complete basis, an arbitrary initial state $\ket{\psi_0}$ may not be able to be expressed by the superpositions of eigenstates, so the QFI may not be zero at EP. Based on Eq. (\ref{f1}) and Eq. (\ref{f2}), we can see that the QFI always tends to zero as EP is approached in both cases of $\mathcal{PT}$-symmetry broken and not, since $\nu_0$ and $\nu_1$ both tend to zero near EP. Therefore, the QFI could be discontinuous at EP for the initial states that are linearly independent of the eigenstates. As shown in Fig. \ref{fig2}(a), we set the initial state as $|0\rangle\langle0|$ at EP ($s=2$) which is linearly independent of the eigenstate, the QFI $\mathcal{F}_{\theta}$ tends to zero as EP is approached, but it suddenly increases to $16$ at EP.

Notice that although the QFI is continuous at EP for the initial states that can be expressed by eigenstates, the channel QFI is still discontinuous at EP. As mentioned above, the QFI tends to zero as EP is approached, but according to Eq. (\ref{max2}), the channel QFI $\mathcal{F}_{\theta,max}$ does not reduce to zero near EP. As shown in Fig. \ref{fig2}(b), in both cases of $\mathcal{PT}-$symmetry broken and not, $\mathcal{F}_{\theta,max}$ does not reduce to zero as EP ($s=2$) is approached, i.e., there still exists some intervals of $\theta$ that $\mathcal{F}_{\theta}\neq0$ near EP. When the system exactly reaches EP, $\mathcal{F}_{\theta,max}$ suddenly reduce to zero. We can see that in Eq. (\ref{f1}) and Eq. (\ref{f2}), the numerator contains $\nu_0$ or $\nu_1$ which gradually tends to zero as EP is approached, hence there is a continuous decrease in $\mathcal{F}_{\theta}$ to zero near EP. However, for $\mathcal{F}_{\theta,max}$, when $\nu_0$ and $\nu_1$ in the numerator tend to zero, there is also an infinitesimal of the same order in the denominator that tends to zero, which makes $\mathcal{F}_{\theta,max}$ not reduce to zero near EP, more details can be found in Appendix C.

\begin{figure}[t]
\centering
\includegraphics[width=0.235\textwidth]{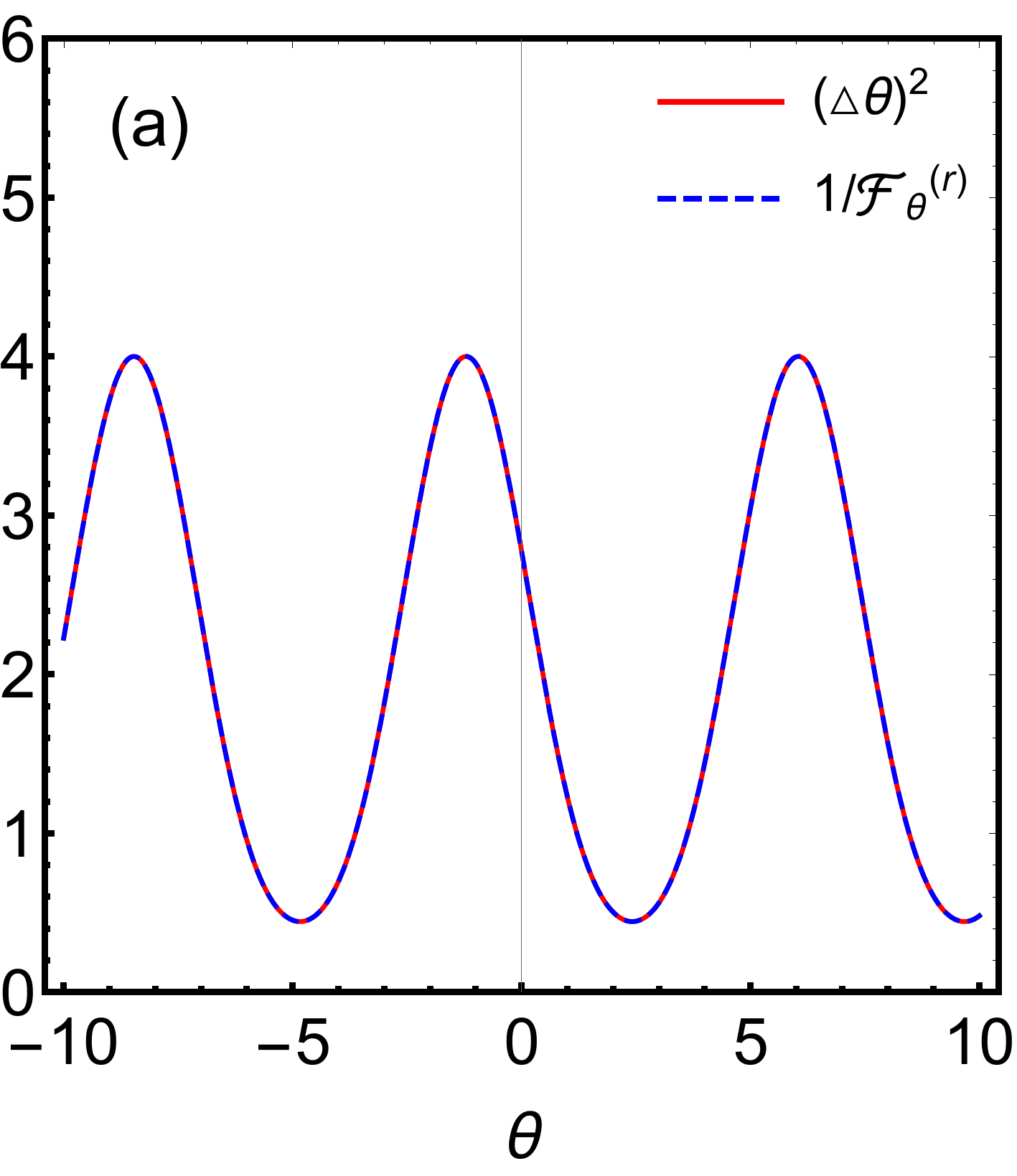}
\includegraphics[width=0.235\textwidth]{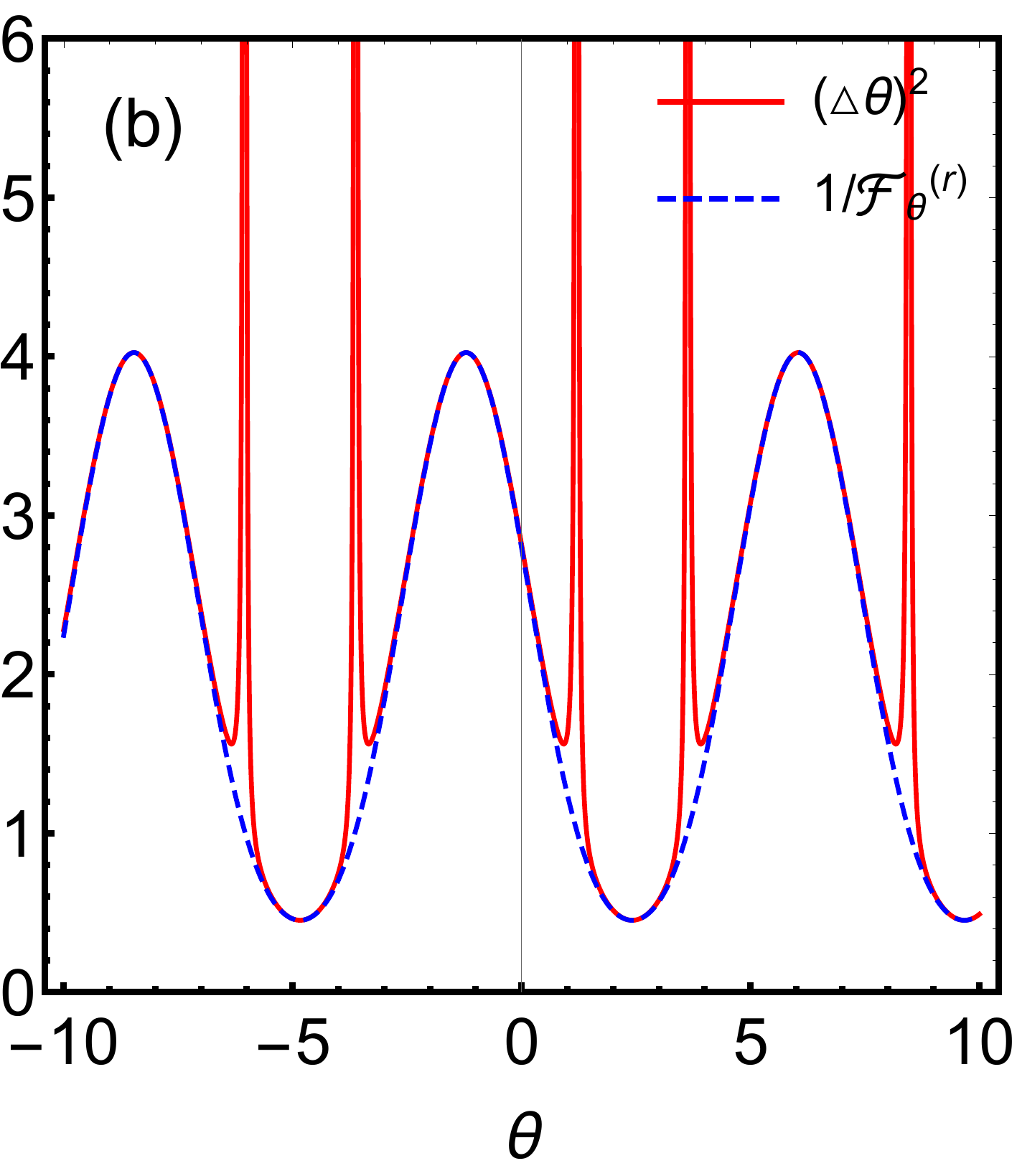}
\includegraphics[width=0.235\textwidth]{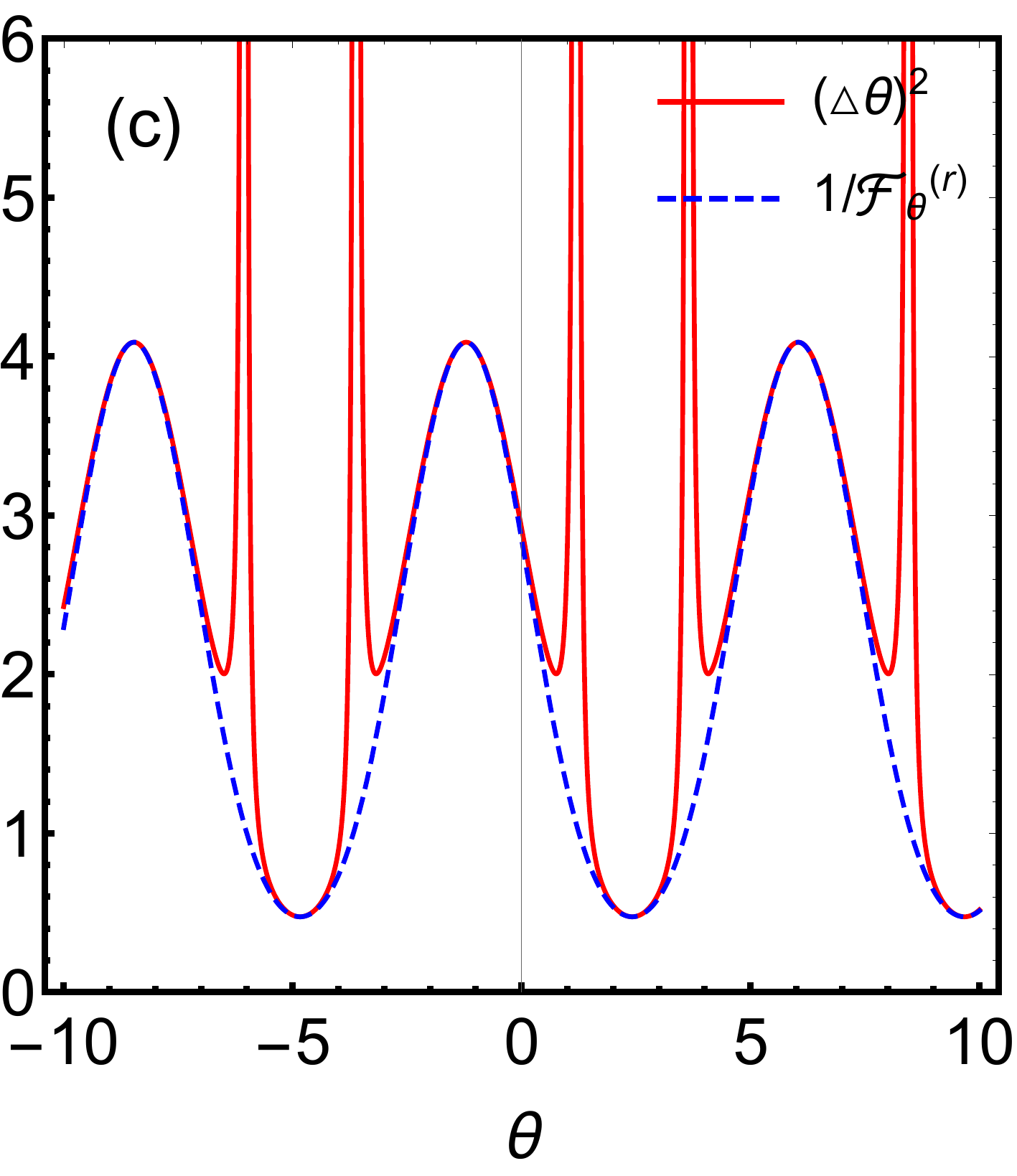}
\includegraphics[width=0.235\textwidth]{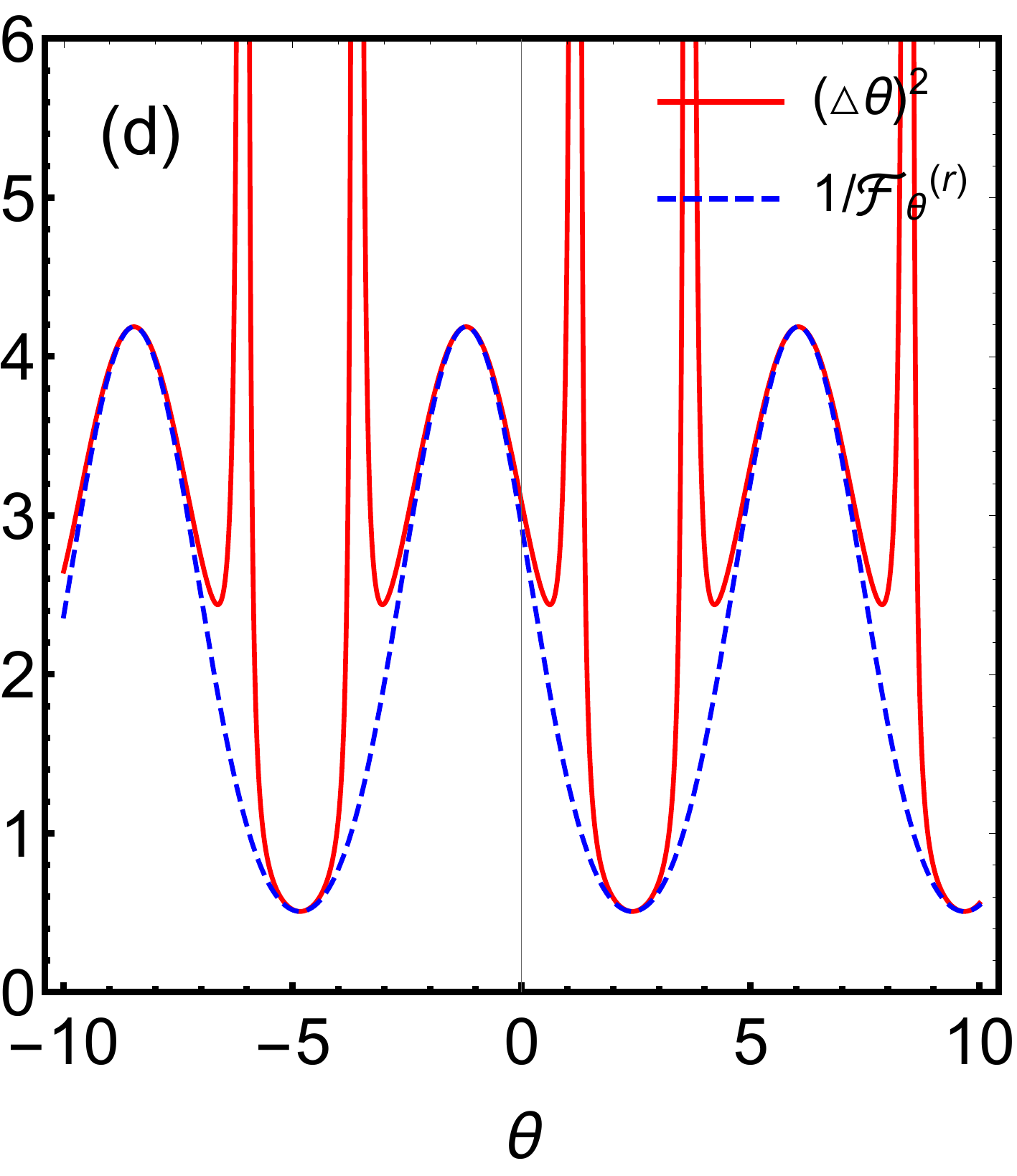}
\caption{The evolution of the variance $(\Delta\theta)^2$ (red solid line) and the reciprocal of QFI $1/\mathcal{F}_{\theta}^{(r)}$ (blue dashed line) as functions of $\theta$, we set the Hamiltonian and the relative phase to remain unchanged, $s=0.5, r=0.25, \varphi=0$. (a) We set the initial state is optimal $m=1.0$. The actual precision $(\Delta\theta)^2$ is equal to $1/\mathcal{F}_{\theta}^{(r)}$, which means the quantum CRB is saturated. (b) We set $m=1.1$, $(\Delta\theta)^2$ is larger than $1/\mathcal{F}_{\theta}^{(r)}$, the precision of the estimation declines. In (c) and (d), we set $m=1.2$ and $m=1.3$ respectively, the precision of the estimation further declines.}\label{optm3}
\end{figure}

On the basis of finding the optimal initial state, we further find an optimal measurement for Hamiltonian $\hat{H}_s$, for simplicity, we set $\omega=\pi/2$. According to Eq. (\ref{a5}), for optimal initial state $\ket{\psi_0}=N(\ket{\lambda_+}+e^{i\varphi}\ket{\lambda_-})$, the corresponding optimal measurement is $|0\rangle\langle0|$, where $\ket{0}=(1,0)^T$, detailed calculation can be found in the Appendix D. Moreover, to verify the condition for optimal measurement that we proposed, we compare the estimation variance $(\Delta\theta)^2$ generated by $|0\rangle\langle0|$ with the reciprocal of QFI $1/\mathcal{F}_{\theta}^{(r)}$. To make the expressions concise, we use $\kappa=r/s$ here, according to error-propagation formula Eq. (\ref{er}), we can obtain the precision of the estimation for arbitrary initial state as follows,
\begin{eqnarray}
(\Delta\theta)^2=p/q,
\end{eqnarray}
where $p=[1+m^2+2m\kappa\sin(2\gamma_0+\beta)]^2[1+m^2-2m\cos2\gamma_0+4m\kappa\sin(2\gamma_0+\beta)](1+m^2+2m\cos2\gamma_0)$, $q=16m^2\nu_0(1-\kappa^2)[2m\kappa+(1+m^2)\sin(2\gamma_0+\beta)]^2$, $\gamma_0=\sqrt{\nu_0}\theta+\varphi/2$ and $\beta$ is determined by $\kappa$ ($\sin\beta=\kappa,\cos\beta=-\sqrt{1-\kappa^2}$).
In Fig. \ref{optm3}(a), we can see that only if $m=1$, $(\Delta\theta)^2$ (the red solid curve) and the $1/\mathcal{F}_{\theta}^{(r)}$ (blue dashed curve) overlap, i.e., the precision of the estimation reaches the quantum CRB. In Fig .\ref{optm3}(b), (c) and (d), the initial state is changed, $(\Delta\theta)^2$ is lager than $1/\mathcal{F}_{\theta}^{(r)}$ in some intervals, the precision of the estimation declines. Thus, $|0\rangle\langle0|$ is exactly the optimal measurement of $\hat{H}_s$ for the optimal initial state.

\section{Discussions and Conclusions}
We simply discuss the method to simulate the $\mathcal{PT}$-symmetric Hamiltonian $\hat{H}_s$ in this section. In short, the expectation value of $\hat{H}_s$ on $\ket{\varphi_\theta}$ can be described with the weak value of the dilated Hamiltonian $\tilde{H}$ on the preselected state $\ket{\tilde{\varphi}_i}$ and postselected state $\ket{\tilde{\varphi}_f}$, i.e. $\bra{\varphi_\theta}\hat{H}_s\ket{\varphi_\theta}_\eta/\langle\varphi_\theta|\varphi_\theta\rangle_\eta=\bra{\tilde{\varphi}_i}\tilde{H}\ket{\tilde{\varphi}_f}/\langle\tilde{\varphi}_i|\tilde{\varphi}_f\rangle$ \cite{rev37}, where $\eta$ is a metric operator, $\ket{\tilde{\varphi}_i}$ and $\ket{\tilde{\varphi}_f}$ are specially dilated from $\ket{\varphi_\theta}$. The right term is exactly the weak value $\langle \tilde{H}\rangle_w$ of dilated Hamiltonian $\tilde{H}$, we can obtain it through weak measurement of $\tilde{H}$. To realize the weak measurement, we need a pointer $\hat{P}$ weakly coupling with $\tilde{H}$, the interaction Hamiltonian can be written as $\hat{H}_{int}=\tilde{H}\otimes\hat{P}$. It has been proved that if we choose $\hat{P}$ as Pauli operator $\hat{\sigma}_z$ and the initial pointer state is $\ket{P_i}=(\ket{+}+\ket{-})/\sqrt{2}$, the real and imaginary parts of the weak value $\langle \tilde{H}\rangle_w$ can be read out by measuring the expectations of Pauli matrices $\hat{\sigma}_y$ and $\hat{\sigma}_z$ respectively on postselected pointer state $\ket{P_f}$ \cite{nev1}.

In conclusion, we have proposed a expression of QFI with general parameter-independent non-Hermitian Hamiltonians. It is four times the variance of non-Hermitian generator on normalized final states, which is quite similar to the QFI of Hermitian Hamiltonians. We also find the condition for optimal measurements, it applies to both Hermitian and non-Hermitian Hamiltonians. Consider a specific $\mathcal{PT}$-symmetric non-Hermitian Hamiltonian as an example, we calculate the QFI and find the optimal measurement. Surprisingly, we find some interesting properties of the QFI of this $\mathcal{PT}$-symmetric Hamiltonian. Considering the application of QFI and $\mathcal{PT}$-symmetry in optics and quantum information, non-Hermitian Hamiltonian QFI is expected to promote the further development of quantum metrology and non-Hermitian physics.

\section*{ACKNOWLEDGMENTS}
This work is  supported by the National Natural Science Foundation of China
(Grant No. 11734015).

\setcounter{equation}{0}
\renewcommand{\theequation}{A\arabic{equation}}
\section*{APPENDIX A: Detailed derivation of QFI for non-Hermitian Hamiltonians}
To clearly generalize the QFI to non-Hermitian systems, we start from classical Fisher information. It is well-known that the Fisher information can be written as
\begin{eqnarray}
F_\theta=\int dx\frac{1}{p(x|\theta)}\Big[\frac{\partial p(x|\theta)}{\partial\theta}\Big]^2
\end{eqnarray}
where $p(x|\theta)=\mathrm{Tr}[\Pi_x\rho_\theta]$, $\rho_\theta=e^{-i\hat{H}\theta}\rho_0e^{i\hat{H}\theta}$ is the final state and $\{\Pi_x\}$ are the elements of a set of positive operator-value measurement (POVM). In non-Hermitian systems, the evolution is non-unitary, as we discussed in the main text, we need to normalized the final state. Thus, we assume that $\tilde{\rho}_\theta$ is the normalized final state:
\begin{eqnarray}
\tilde{\rho}_\theta=\frac{e^{-iH\theta}\rho_0e^{i\hat{H}^\dagger\theta}}{\mathrm{Tr}[e^{-iH\theta}\rho_0e^{i\hat{H}^\dagger\theta}]}=\frac{\rho_\theta}{K_\theta},
\end{eqnarray}
then we have $p(x|\theta)=\mathrm{Tr}[\Pi_x\tilde{\rho}_\theta]$. Noticed that even if the state $\rho_0$ evolves in non-Hermitian system, the final state $\tilde{\rho}_\theta$ is still Hermitian, and obviously we can still find a Hermitian Symmetric Logarithmic Derivative operator $L_\theta$ satisfies the equation $(L_\theta\tilde{\rho}_\theta+\tilde{\rho}_\theta L_\theta)/2=\partial_\theta\tilde{\rho}_\theta$, and in the eigenbasis of $\tilde{\rho}_{\theta}=\sum_{n}c_n|\psi_n\rangle\langle\psi_n|$, $L_\theta$ can be written as follows
\begin{eqnarray}
L_\theta=2\sum_{n,m}\frac{\bra{\psi_n}\partial_{\theta}\tilde{\rho}_{\theta}\ket{\psi_m}}{c_n+c_m}|\psi_n\rangle\langle\psi_m|, \label{SLD}
\end{eqnarray}
where $c_n+c_m$ is required not to be zero. Obviously this expression satisfies the equation $(L_\theta\tilde{\rho}_\theta+\tilde{\rho}_\theta L_\theta)/2=\partial_\theta\tilde{\rho}_\theta$. Then we have
\begin{eqnarray}
\partial_\theta p(x|\theta)&=&\partial_\theta\mathrm{Tr}[\Pi_x\tilde{\rho}_\theta]=\mathrm{Tr}[\Pi_x\partial_\theta\tilde{\rho}_\theta]\nonumber\\
&=&\frac{\mathrm{Tr}[\Pi_xL_\theta\tilde{\rho}_\theta]+\mathrm{Tr}[\Pi_x\tilde{\rho}_\theta L_\theta]}{2}\nonumber\\
&=&\frac{\mathrm{Tr}[\Pi_xL_\theta\tilde{\rho}_\theta]+\mathrm{Tr}[(\Pi_x\tilde{\rho}_\theta L_\theta)^\dagger]^\ast}{2}\nonumber\\
&=&\frac{\mathrm{Tr}[\Pi_xL_\theta\tilde{\rho}_\theta]+\mathrm{Tr}[L_\theta\tilde{\rho}_\theta\Pi_x]^\ast}{2}\nonumber\\
&=&\mathrm{Re}\{\mathrm{Tr}[\Pi_xL_\theta\tilde{\rho}_\theta]\}.
\end{eqnarray}
With this result, the Fisher information can be further written as
\begin{eqnarray}
F_\theta=\int dx\frac{\mathrm{Re}\{\mathrm{Tr}[\Pi_xL_\theta\tilde{\rho}_\theta]\}^2}{\mathrm{Tr}[\Pi_x\tilde{\rho}_\theta]}.
\end{eqnarray}

As it known to us, the quantum Fisher information is the maximum Fisher information over different valid quantum measurements, so we amplify the Fisher information as
\begin{eqnarray}
F_\theta&=&\int dx\frac{\mathrm{Re}\{\mathrm{Tr}[\Pi_xL_\theta\tilde{\rho}_\theta]\}^2}{\mathrm{Tr}[\Pi_x\tilde{\rho}_\theta]}\leq\int dx\frac{\big|\mathrm{Tr}[\Pi_xL_\theta\tilde{\rho}_\theta]\big|^2}{\mathrm{Tr}[\Pi_x\tilde{\rho}_\theta]}\nonumber\\
&=&\int dx\frac{\Big|\mathrm{Tr}\big[(\sqrt{\tilde{\rho_\theta}}\sqrt{\Pi_x})(\sqrt{\Pi_x}L_\theta\sqrt{\tilde{\rho}_\theta})\big]\Big|^2}{\mathrm{Tr}[\Pi_x\tilde{\rho}_\theta]}.
\end{eqnarray}
According to the Schwartz inequality: $|\mathrm{Tr}[AB]|^2\leq\mathrm{Tr}[A^\dagger A]\mathrm{Tr}[B^\dagger B]$, we have
\begin{eqnarray}
F_\theta&\leq&\int dx\frac{\Big|\mathrm{Tr}\big[(\sqrt{\tilde{\rho}_\theta}\sqrt{\Pi_x})(\sqrt{\Pi_x}L_\theta\sqrt{\tilde{\rho}_\theta})\big]\Big|^2}{\mathrm{Tr}[\Pi_x\tilde{\rho}_\theta]}\nonumber\\
&\leq&\int dx\frac{\mathrm{Tr}\big[\tilde{\rho}_\theta\Pi_x\big]\mathrm{Tr}\big[\Pi_xL_\theta\tilde{\rho}_\theta L_\theta\big]}{\mathrm{Tr}[\Pi_x\tilde{\rho}_\theta]}\nonumber\\
&=&\int dx\mathrm{Tr}\big[\Pi_xL_\theta\tilde{\rho}_\theta L_\theta\big]\nonumber\\
&=&\mathrm{Tr}\big[(\int dx\Pi_x)L_\theta\tilde{\rho}_\theta L_\theta\big]=\mathrm{Tr}\big[\tilde{\rho}_\theta L_\theta^2\big].
\end{eqnarray}
According to the above chain of inequalities, we obtain the quantum Fisher information
\begin{eqnarray}
\mathcal{F}_\theta=\mathrm{Tr}\big[\tilde{\rho}_\theta L_\theta^2\big]=\mathrm{Tr}\big[(\partial_\theta\tilde{\rho}_\theta) L_\theta\big], \label{QFIL}
\end{eqnarray}
which is the upper bound of Fisher information. We can see that it is the same as the form in Hermitian systems. Substitute Eq. (\ref{SLD}) into the last expression of Eq. (\ref{QFIL}), we obtain
\begin{eqnarray}
\mathcal{F}_\theta&=&\mathrm{Tr}\big[\partial_\theta\tilde{\rho}_\theta(2\sum_{n,m}\frac{\bra{\psi_n}\partial_{\theta}\tilde{\rho}_{\theta}\ket{\psi_m}}{c_n+c_m}|\psi_n\rangle\langle\psi_m|)\big]\nonumber\\
&=&2\sum_{n,m}\frac{\bra{\psi_n}\partial_{\theta}\tilde{\rho}_{\theta}\ket{\psi_m}}{c_n+c_m}\mathrm{Tr}\big[\partial_\theta\tilde{\rho}_\theta|\psi_n\rangle\langle\psi_m|\big]\nonumber\\
&=&2\sum_{n,m}\frac{|\bra{\psi_n}\partial_{\theta}\tilde{\rho}_{\theta}\ket{\psi_m}|^2}{c_n+c_m}.\label{App1}
\end{eqnarray}

For a generic family of pure states, the normalized final state is
\begin{eqnarray}
\tilde{\rho}_\theta=|\varphi_\theta\rangle\langle\varphi_\theta|=\frac{|\psi_\theta\rangle\langle\psi_\theta|}{K_\theta}.
\end{eqnarray}
where $K_{\theta}=\langle\psi_\theta|\psi_\theta\rangle$ is the normalization coefficient. According to Eq. (\ref{App1}), we can obtain
\begin{eqnarray}
\mathcal{F}_\theta&=&2[\frac{|\langle\varphi_\theta|\partial_\theta\varphi_\theta\rangle+\langle\partial_\theta\varphi_\theta|\varphi_\theta\rangle|^2}{2}+2\sum_{i}|\langle\partial_\theta\varphi_\theta|i\rangle|^2]\nonumber\\
&=&2[\frac{|\partial_\theta(\langle\varphi_\theta|\varphi_\theta\rangle)|^2}{2}+2\sum_{i}\langle\partial_\theta\varphi_\theta|i\rangle\langle i|\partial_\theta\varphi_\theta\rangle]\nonumber\\
&=&4\langle\partial_\theta\varphi_\theta|(\sum_{i}|i\rangle\langle i|)|\partial_\theta\varphi_\theta\rangle\nonumber\\
&=&4\langle\partial_\theta\varphi_\theta|(\mathrm{I}-|\varphi_\theta\rangle\langle\varphi_\theta|)|\partial_\theta\varphi_\theta\rangle\nonumber\\
&=&4(\langle\partial_{\theta}\varphi_{\theta}|\partial_{\theta}\varphi_{\theta}\rangle-|\langle\partial_{\theta}\varphi_{\theta}|\varphi_{\theta}\rangle|^{2}),\label{A2}
\end{eqnarray}
\begin{eqnarray}
(L_\theta\rho_\theta+\rho_\theta L_\theta)/2=\partial_\theta\rho_\theta,
\end{eqnarray}
where $\ket{\partial_\theta\varphi_\theta}=\partial_\theta|\varphi_\theta\rangle$. And then, we further expand $|\partial_{\theta}\varphi_{\theta}\rangle$ with $\hat{U}_{\theta}$ and $K_{\theta}$, we have $|\partial_{\theta}\varphi_{\theta}\rangle=\partial_{\theta}(\hat{U}_{\theta}/\sqrt{K_{\theta}})|\psi_{0}\rangle$. Thus, we can obtain the first term:
\begin{eqnarray}
&&\langle\partial_{\theta}\varphi_{\theta}|\partial_{\theta}\varphi_{\theta}\rangle\nonumber\\
&=&\langle\psi_{0}|\frac{\sqrt{K_{\theta}}(\partial_{\theta}\hat{U}_{\theta}^{\dagger})-\frac{(\partial_{\theta}K_{\theta})}{2\sqrt{K_{\theta}}}\hat{U}_{ \theta}^{\dagger}}{K_{\theta}}\nonumber\\
&&\ \ \ \ \ \  \cdot\frac{\sqrt{K_{\theta}}(\partial_{\theta}\hat{U}_{\theta})-\frac{(\partial_{\theta}K_{\theta})}{2\sqrt{K_{\theta}}}\hat{U}_ {\theta}}{K_{\theta}}|\psi_{0}\rangle\nonumber\\
&=&\langle \hat{H}^{\dagger}\hat{H}\rangle_\theta-\frac{i(\partial_{\theta}K_{\theta})}{2K_{\theta}}(\langle \hat{H}^{\dagger}\rangle_\theta-\langle \hat{H}\rangle_\theta)+\frac{(\partial_{\theta}K_{\theta})^{2}}{4K_{\theta}^{2}},\nonumber\\
\end{eqnarray}
where $\hat{U}_\theta=e^{-i\hat{H}\theta}$ and $\partial_\theta\hat{U}_\theta=-i\hat{U}_\theta\hat{H}$, note that $\hat{H}$ commutes with $\hat{U}_\theta$. Here, we define $\langle \hat{X}\rangle_\theta$ as the expectation of operator $\hat{X}$ on normalized final state ($\langle \hat{X}\rangle_\theta=\bra{\varphi_{\theta}}\hat{X}\ket{\varphi_{\theta}}$). In the same way, we can also obtain the expression for the second term as follows:
\begin{eqnarray}
&&|\langle\partial_{\theta}\varphi_\theta|\varphi_{\theta}\rangle|^{2}\nonumber\\
&=&|\bra{\psi_{0}}\frac{\sqrt{K_{\theta}}(\partial_{\theta}\hat{U}_{\theta}^{\dagger})-\frac{(\partial_{\theta}K_{\theta})}{2\sqrt{K_{\theta}}}\hat{U}_{ \theta}^{\dagger}}{K_{\theta}}\frac{\hat{U}_{\theta}}{\sqrt{K_{\theta}}}\ket{\psi_{0}}|^{2}\nonumber\\
&=&|\bra{\psi_{0}}\frac{i\hat{U}_{\theta}^{\dagger}\hat{H}^{\dagger}\hat{U}_{\theta}-\frac{(\partial_{\theta}K_{\theta})}{2K_{\theta}}\hat{U}_{\theta}^{\dagger}\hat{U}_{\theta}}{K_{\theta}}\ket{\psi_{0}}|^{2}\nonumber\\
&=&|i\langle \hat{H}^{\dagger}\rangle_\theta-\frac{(\partial_{\theta}K_{\theta})}{2K_{\theta}}|^2\nonumber\\
&=&\langle \hat{H}^{\dagger}\rangle_\theta\langle \hat{H}\rangle_\theta-\frac{i(\partial_{\theta}K_{\theta})}{2K_{\theta}}(\langle \hat{H}^{\dagger}\rangle_\theta-\langle \hat{H}\rangle_\theta)+\frac{(\partial_{\theta}K_{\theta})^{2}}{4K_{\theta}^{2}}.\nonumber\\
\end{eqnarray}
Substituting these two results back into Eq. (\ref{A2}), we work out the concise and intuitive expression of QFI,
\begin{eqnarray}
\mathcal{F}_{\theta}=4(\langle \hat{H}^{\dagger}\hat{H}\rangle_\theta-\langle \hat{H}^{\dagger}\rangle_\theta\langle \hat{H}\rangle_\theta),\label{A3}
\end{eqnarray}
If the Hamiltonian is Hermitian, this result returns to $\mathcal{F}_{\theta}=4\bra{\psi_0}(\Delta\hat{H})^2\ket{\psi_0}$.

\setcounter{equation}{0}
\renewcommand{\theequation}{B\arabic{equation}}
\section*{APPENDIX B: Proof of the condition for optimal measurements}
Suppose we have a Hermitian operator $\hat{A}$ as a measurement, according to the error-propagation formula \cite{ero1,ero2}, we have
\begin{eqnarray}
(\Delta\theta)^2&=&\frac{(\Delta \hat{A})^2}{n|\partial_\theta \langle \hat{A}\rangle_\theta|^2} \nonumber\\
&=&\frac{(\Delta\hat{A})^2}{n|\bra{\partial_{\theta}\varphi_{\theta}}\hat{A}\ket{\varphi_{\theta}}+\bra{\varphi_\theta}\hat{A}\ket{\partial_\theta\varphi_{\theta}}|^2},\label{A4}
\end{eqnarray}
where $\ket{\varphi_\theta}$ is the normalized final state.

In the case of Non-Hermitian Hamiltonians, let $Q=\partial_\theta \langle \hat{A}\rangle_\theta=\bra{\partial_{\theta}\varphi_{\theta}}\hat{A}\ket{\varphi_{\theta}}+\bra{\varphi_\theta}\hat{A}\ket{\partial_\theta\varphi_{\theta}}$ and expand this expression, we have
\begin{eqnarray}\label{Bq}
Q&=&\bra{\psi_0}\partial_\theta(\frac{\hat{U}_\theta^\dagger}{\sqrt{K_\theta}})\hat{A}\ket{\varphi_\theta}+\bra{\varphi_\theta}\hat{A}\partial_\theta(\frac{\hat{U}_\theta}{\sqrt{K_\theta}})\ket{\psi_0}\nonumber\\
&=&\bra{\psi_0}\frac{\sqrt{K_{\theta}}\hat{U}_{\theta}^\dagger(i\hat{H}^{\dagger})-\frac{\partial_\theta K_\theta}{2\sqrt{K_\theta}}\hat{U}_\theta^\dagger}{K_{\theta}}\hat{A}\ket{\varphi_\theta}+\bar{Z}\nonumber\\
&=&\bra{\varphi_\theta}(i\hat{H}^{\dagger}-\frac{\partial_\theta K_\theta}{2K_\theta})\hat{A}\ket{\varphi_\theta}-\bra{\varphi_\theta}\hat{A}(i\hat{H}+\frac{\partial_\theta K_\theta}{2K_\theta})\ket{\varphi_\theta}\nonumber\\
&=&\bra{\varphi_\theta}i(\hat{H}^{\dagger} \hat{A}-\hat{A}\hat{H})\ket{\varphi_\theta}-\frac{\partial_\theta K_\theta}{K_\theta}\bra{\varphi_\theta}\hat{A}\ket{\varphi_\theta}\nonumber\\
&=&i(\langle \hat{H}^{\dagger} \hat{A}\rangle_\theta-\langle \hat{A}\hat{H}\rangle_\theta)-\frac{\partial_\theta K_\theta}{K_\theta}\langle \hat{A}\rangle_\theta,
\end{eqnarray}
where $  \bar{Z}=\bra{\varphi_\theta}\hat{A}\partial_\theta(\hat{U}_\theta/\sqrt{K_\theta})\ket{\psi_0}$, $\partial_\theta(\hat{U}_\theta/\sqrt{K_\theta})=[\sqrt{K_{\theta}}\hat{U}_{\theta}(-i\hat{H})-(\partial_\theta K_\theta/2\sqrt{K_\theta})\hat{U}_\theta]/K_\theta$, and $K_{\theta}=\bra{\psi_\theta}\psi_\theta\rangle$ is the normalization coefficient of final state, then $(\partial_\theta K_\theta)/K_\theta$ can be written as
\begin{eqnarray}
\frac{\partial_\theta K_\theta}{K_\theta}&=&\frac{\partial_\theta(\bra{\psi_0}\hat{U}_\theta^\dagger \hat{U}_\theta\ket{\psi_0})}{K_\theta}\nonumber\\
&=&\frac{\bra{\psi_0}\hat{U}_\theta^\dagger i\hat{H}^{\dagger} \hat{U}_\theta\ket{\psi_0}+\bra{\psi_0}\hat{U}_\theta^\dagger(-i\hat{H}) \hat{U}_\theta\ket{\psi_0}}{K_\theta}\nonumber\\
&=&\bra{\varphi_\theta}i(\hat{H}^{\dagger}-\hat{H})\ket{\varphi_\theta}\nonumber\\
&=&i(\langle \hat{H}^{\dagger}\rangle_\theta-\langle \hat{H}\rangle_\theta). \label{pk}
\end{eqnarray}
Substitute the result into Eq. (\ref{Bq}), we have
\begin{eqnarray}
Q=i[(\langle\hat{H}^{\dagger}\hat{A}\rangle_\theta-\langle \hat{A}\hat{H}\rangle_\theta)-(\langle \hat{H}^{\dagger}\rangle_\theta-\langle \hat{H}\rangle_\theta)\hat{A}\rangle_\theta].\nonumber\\
\end{eqnarray}
Then $|Q|^2$ can be written as
\begin{eqnarray}
|Q|^2&=&-(\langle \hat{H}^{\dagger} \hat{A}\rangle_\theta-\langle\hat{A}\hat{H}\rangle_\theta)^2-(\langle \hat{H}^{\dagger}\rangle_\theta-\langle \hat{H}\rangle_\theta)^2\langle \hat{A}\rangle_\theta^2\nonumber\\
&&+2(\langle \hat{H}^{\dagger}\rangle_\theta-\langle \hat{H}\rangle_\theta)(\langle \hat{H}^{\dagger} \hat{A}\rangle_\theta-\langle\hat{A}\hat{H}\rangle_\theta)\langle \hat{A}\rangle_\theta\nonumber\\
&=&-[(\langle \hat{H}^{\dagger} \hat{A}\rangle_\theta+\langle\hat{A}\hat{H}\rangle_\theta)^2+(\langle \hat{H}^{\dagger}\rangle_\theta+\langle \hat{H}\rangle_\theta)^2\langle \hat{A}\rangle_\theta^2\nonumber\\
&&-2(\langle \hat{H}^{\dagger}\rangle_\theta+\langle \hat{H}\rangle_\theta)(\langle \hat{H}^{\dagger} \hat{A}\rangle_\theta+\langle\hat{A}\hat{H}\rangle_\theta)\langle \hat{A}\rangle_\theta]\nonumber\\
&&+4[\langle \hat{H}^{\dagger} \hat{A}\rangle_\theta\langle\hat{A}\hat{H}\rangle_\theta+\langle \hat{H}^{\dagger}\rangle_\theta\langle \hat{H}\rangle_\theta\langle \hat{A}\rangle_\theta^2-\nonumber\\
&&\langle \hat{H}^{\dagger} \hat{A}\rangle_\theta\langle \hat{H}\rangle_\theta\langle \hat{A}\rangle_\theta-\langle\hat{A}\hat{H}\rangle_\theta\langle \hat{H}^{\dagger}\rangle_\theta\langle \hat{A}\rangle_\theta].\label{B1}
\end{eqnarray}
According to the non-Hermitian uncertainty relationship $(\Delta\hat{A})^2(\Delta\hat{B})^2\geq|\langle\hat{A}^{\dagger}\hat{B}\rangle-\langle\hat{A}^{\dagger}\rangle\langle\hat{B}\rangle|^2$ \cite{ur1,ur2,ur3,rev35,zxev35} which was first proposed by Pati in \cite{ur1}, we have
\begin{eqnarray}\label{B2}
(\Delta\hat{H})^2(\Delta\hat{A})^2&\geq&|\langle \hat{H}^{\dagger}\hat{A}\rangle_\theta-\langle \hat{H}^{\dagger}\rangle\langle \hat{A}\rangle_\theta|^2\nonumber\\
&=&\langle \hat{H}^{\dagger} \hat{A}\rangle_\theta\langle\hat{A}\hat{H}\rangle_\theta+\langle \hat{H}^{\dagger}\rangle_\theta\langle \hat{H}\rangle_\theta\langle \hat{A}\rangle_\theta^2\nonumber\\
&&-\langle \hat{H}^{\dagger} \hat{A}\rangle_\theta\langle \hat{H}\rangle_\theta\langle \hat{A}\rangle_\theta-\langle\hat{A}\hat{H}\rangle_\theta\langle \hat{H}^{\dagger}\rangle_\theta\langle \hat{A}\rangle_\theta,\nonumber\\
\end{eqnarray}
where $(\Delta\hat{H})^2=\langle \hat{H}^{\dagger}\hat{H}\rangle_\theta-\langle \hat{H}^{\dagger}\rangle_\theta\langle \hat{H}\rangle_\theta$. The expression in the last brackets of Eq. (\ref{B1}) is exactly the same as $(\Delta\hat{H})^2(\Delta\hat{A})^2$. Then, we can derive an inequality as follow,
\begin{eqnarray}\label{B3}
|Q|^2&\leq&4(\Delta\hat{H})^2(\Delta\hat{A})^2-[(\langle \hat{H}^{\dagger} \hat{A}\rangle_\theta+\langle\hat{A}\hat{H}\rangle_\theta)-\nonumber\\
&&(\langle \hat{H}^{\dagger}\rangle_\theta+\langle \hat{H}\rangle_\theta)\langle \hat{A}\rangle_\theta]^2.
\end{eqnarray}
Here, we define an operator $\delta \hat{M}=\hat{M}-\langle\hat{M}\rangle_{\theta}$, then we have
\begin{eqnarray}
&&\ket{f}=\delta \hat{H}\ket{\varphi_\theta}=\hat{H}\ket{\varphi_\theta}-\langle \hat{H}\rangle_{\theta}\ket{\varphi_\theta},\label{B5}\\
&&\ket{g}=\delta \hat{A}\ket{\varphi_\theta}=\hat{A}\ket{\varphi_\theta}-\langle \hat{A}\rangle_{\theta}\ket{\varphi_\theta},\label{B6}\\
&&\bra{f}g\rangle=\langle \hat{H}^{\dagger} \hat{A}\rangle_\theta-\langle \hat{H}^{\dagger}_\theta\rangle\langle \hat{A}\rangle_\theta.
\end{eqnarray}
Then the inequality Eq. (\ref{B3}) can be written as
\begin{eqnarray}
|Q|^2&\leq&4(\Delta\hat{H})^2(\Delta\hat{A})^2-[(\langle \hat{H}^{\dagger} \hat{A}\rangle_\theta-\langle \hat{H}^{\dagger}\rangle_\theta\langle \hat{A}\rangle_\theta)+\nonumber\\
&&(\langle\hat{A}\hat{H}\rangle_\theta-\langle \hat{H}\rangle_\theta\langle \hat{A}\rangle_\theta)]^2\nonumber\\
&=&4(\Delta\hat{H})^2(\Delta\hat{A})^2-(\bra{f}g\rangle+\bra{g}f\rangle)\nonumber\\
&\leq&4(\Delta\hat{H})^2(\Delta\hat{A})^2.
\end{eqnarray}
According to the Cauchy-Schwarz inequality, the first inequality is saturated only if $\ket{f}=C\ket{g}$, where $C$ is a constant. Furthermore, if $C$ is an imaginary number, the second inequality will be saturated. Then we have
\begin{eqnarray}
(\Delta\theta)^2&=&\frac{(\Delta\hat{A})^2}{n|\partial_\theta \langle \hat{A}\rangle_\theta|^2}\geq\frac{1}{4(\Delta\hat{H})^2}=\frac{1}{n\mathcal{F}_{\theta}}.
\end{eqnarray}
Now we obtain the condition of optimal measurements as follows,
\begin{eqnarray}
\ket{f}=iC\ket{g},\label{B4}
\end{eqnarray}
where $C$ is a real number. Actually, this relationship is also useful for Hermitian Hamiltonians, which has already been obtained in Ref. \cite{rev36}.

As for Hermitian Hamiltonians, the error-propagation formula can be expanded as follows:
\begin{eqnarray}
(\Delta\theta)^2&=&\frac{(\Delta\hat{A})^2}{n|\partial_\theta\langle\hat{A}\rangle_\theta|^2} \nonumber\\
&=&\frac{(\Delta\hat{A})^2}{n|\bra{\partial_{\theta}\varphi_{\theta}}\hat{A}\ket{\varphi_{\theta}}+\bra{\varphi_\theta}\hat{A}\ket{\partial_\theta\varphi_{\theta}}|^2}\nonumber\\
&=&\frac{(\Delta\hat{A})^2}{n|\langle\hat{H}\hat{A}-\hat{A}\hat{H}\rangle|^2}.
\end{eqnarray}
According to Heisenberg uncertainty relation, we have $|\langle\hat{H}A-A\hat{H}\rangle|^2\leq4(\Delta\hat{A})^2(\Delta\hat{H})^2$. Therefore, the variance of $\theta$ is limited,
\begin{eqnarray}
(\Delta\theta)^2\geq\frac{1}{4n(\Delta\hat{H})^2}=\frac{1}{n\mathcal{F}_{\theta}}.
\end{eqnarray}
The condition that saturates the inequality is also $\ket{f}=iC\ket{g}$. Hofmann has proposed this condition in 2009 \cite{rev36}.

We next prove that Symmetric Logarithmic Derivative (SLD) satisfies Eq. (\ref{B4}). As we mentioned above, SLD $L_\theta$ satisfies the equation $(L_\theta\tilde{\rho}_\theta+\tilde{\rho}_\theta L_\theta)/2=\partial_\theta\tilde{\rho}_\theta$. And we have
\begin{eqnarray}
&\partial_\theta(\tilde{\rho}^2_\theta)=(\partial_\theta\tilde{\rho}_\theta)\tilde{\rho}_\theta+\tilde{\rho}_\theta(\partial_\theta\tilde{\rho}_\theta)&\\
&\partial_\theta(\tilde{\rho}^2_\theta)=\partial_\theta(\tilde{\rho}_\theta).&
\end{eqnarray}
Obviously, SLD is $2\partial_\theta\tilde{\rho}_\theta$ for pure state model, i.e.,
\begin{eqnarray}
L_\theta=2(|\partial_\theta\varphi_\theta\rangle\langle\varphi_\theta|+|\varphi_\theta\rangle\langle\partial_\theta\varphi_\theta|).
\end{eqnarray}
We further expand $L_\theta$ as follows,
\begin{eqnarray}
L_\theta&=&2(|\partial_\theta\varphi_\theta\rangle\langle\varphi_\theta|+|\varphi_\theta\rangle\langle\partial_\theta\varphi_\theta|)\nonumber\\
&=&2\Big[(-i\hat{H}|\varphi_\theta\rangle\langle\varphi_\theta|+|\varphi_\theta\rangle\langle\varphi_\theta|i\hat{H}^\dagger)-\frac{\partial_\theta K_\theta}{K_\theta}|\varphi_\theta\rangle\langle\varphi_\theta|\Big].\nonumber\\
\end{eqnarray}
According to Eq. (\ref{pk}), we have
\begin{eqnarray} \label{SLDexpd}
L_\theta&=&-2i\Big[(\hat{H}-\langle\hat{H}\rangle_\theta)|\varphi_\theta\rangle\langle\varphi_\theta|-|\varphi_\theta\rangle\langle\varphi_\theta|(\hat{H}^\dagger-\langle\hat{H}^{\dagger}\rangle_\theta)\Big]\nonumber\\
&=&-2i\Big[\delta\hat{H}|\varphi_\theta\rangle\langle\varphi_\theta|-|\varphi_\theta\rangle\langle\varphi_\theta|\delta\hat{H}^\dagger\Big].
\end{eqnarray}
Therefore, we obtain
\begin{eqnarray} \label{SLDret}
\ket{g}&=&\delta L_\theta\ket{\varphi_\theta}\nonumber\\
&=&-2i\Big[\big(\delta\hat{H}|\varphi_\theta\rangle\langle\varphi_\theta|-|\varphi_\theta\rangle\langle\varphi_\theta|\delta\hat{H}^\dagger\big)\nonumber\\
&&\ \ \ \ \ \ -\big(\langle\delta\hat{H}\rangle_\theta-\langle\delta\hat{H}^\dagger\rangle_\theta\big)\Big]\ket{\varphi_\theta}\nonumber\\
&=&-2i\delta\hat{H}\ket{\varphi_\theta}=-2i\ket{f}.
\end{eqnarray}
Obviously, it satisfies Eq. (\ref{B4}), so SLD satisfies the condition for optimal measurement. However, not only Eq. (\ref{SLDexpd}) satisfies the condition, we can also find other operators that satisfies the condition. We just need to add a term on Eq. (\ref{SLDexpd}), whose average value $\bra{\varphi_\theta}\bullet\ket{\varphi_\theta}$ and the result acting on $\ket{\varphi_\theta}$ are $0$, then the result of Eq. (\ref{SLDret}) will not be changed. For example,
\begin{eqnarray}
L'_\theta&=&-2i\Big[\delta\hat{H}|\varphi_\theta\rangle\langle\varphi_\theta|-|\varphi_\theta\rangle\langle\varphi_\theta|\delta\hat{H}^\dagger\Big]\nonumber\\
&& +C\delta\hat{M}|\varphi_\theta\rangle\langle\varphi_\theta|\delta\hat{M}^\dagger,\nonumber
\end{eqnarray}
where $C$ is a real number and $\hat{M}$ is an arbitrary operator. Obviously, $L'_\theta$ also satisfies the condition for optimal measurement and the equation $(L'_\theta\tilde{\rho}_\theta+\tilde{\rho}_\theta L'_\theta)/2=\partial_\theta\tilde{\rho}_\theta$. Thus, we can find more optimal measurements with our condition.

\setcounter{equation}{0}
\renewcommand{\theequation}{C\arabic{equation}}
\section*{APPENDIX C:  QFI of a specific $\mathcal{PT}$-symmetric non-Hermitian Hamiltonian}

Here, we provide the detailed calculations of the QFI for a specific $\mathcal{PT}$-symmetric Hamiltonian based on Eq. (\ref{A3}). For a given two level $\mathcal{PT}$-symmetric Hamiltonian:
\begin{eqnarray}
\hat{H}_s=\begin{pmatrix} re^{i\omega} & s \\ s & re^{-i\omega}  \end{pmatrix}.
\end{eqnarray}
where $s>0$, $r$ and $\omega$ are real. We can see that the eigenvalues $\lambda_{\pm}=\mu\pm\sqrt{\nu_0}=r\cos{\omega}\pm\sqrt{s^2-r^2\sin^2{\omega}}$ are still real when the $\mathcal{PT}$-symmetry is not broken, i.e. $s^2>r^2\sin^2{\omega}$. And the normalized eigenstates are
\begin{eqnarray}
\ket{\lambda_+}=\frac{1}{\sqrt{2}}\begin{pmatrix} e^{i\alpha/2} \\ e^{-i\alpha/2} \end{pmatrix},\quad
\ket{\lambda_-}=\frac{i}{\sqrt{2}}\begin{pmatrix} e^{-i\alpha/2} \\ -e^{i\alpha/2} \end{pmatrix}.\nonumber
\end{eqnarray}
where $\sin{\alpha}=(r/s)\sin{\omega}$. As discussed above, it is necessary to derive the normalized coefficient $K_\theta$. For an arbitrary initial state $\ket{\psi_0}=N(\ket{\lambda_+}+me^{i\varphi}\ket{\lambda_-})$ written in $\hat{H}_s$-present,
we are able to work out the normalized coefficient $K_\theta$ as follows:
\begin{eqnarray}
K_{\theta}^{(r)}&=&\bra{\psi_{0}}\hat{U}_{\theta}^{\dagger}\hat{U}_{\theta}\ket{\psi_{0}}\nonumber\\
&=&N^{2}\{e^{i(\lambda_{+}^{\ast}-\lambda_{+})\theta}+m^{2}e^{i(\lambda_{-}^{\ast}-\lambda_{-})\theta}+\nonumber\\
&&2m\textbf{Re}[e^{i[(\lambda_{+}^{\ast}-\lambda_{-})\theta+\varphi]}\lambda_{+}^{\ast}\lambda_{-}\langle\lambda_{+}|\lambda_{-}\rangle]\}\nonumber\\
&=&N^{2}\{1+m^{2}+2m\sin\alpha\cos\gamma\},
\end{eqnarray}
where $\gamma=2\sqrt{\nu_0}\theta+\varphi$. Let $N_\theta^{(r)}=N^{2}/K_{\theta}^{(r)}$, then $\langle \hat{H}_s^{\dagger}\hat{H}_s\rangle$ and $\langle \hat{H}_s\rangle$ can be written as
\begin{eqnarray}
\langle \hat{H}_s^{\dagger}\hat{H}_s\rangle_\theta&=&N_{\theta}^{(r)}(|\lambda_{+}|^{2}+m^{2}|\lambda_{-}|^2+\nonumber\\
&&2\textbf{Re}[m\lambda_{+}^{\ast}\lambda_{-}e^{i[(\lambda_{+}^{\ast}-\lambda_{-})\theta+\varphi]}\langle\lambda_{+}|\lambda_{-}\rangle])\nonumber\\
&=&N_{\theta}^{(r)}[(r\cos\omega+\sqrt{\nu_0})^2+m^2(r\cos\omega-\sqrt\nu_0)^2\nonumber\\
&&+2m(r^2\cos^2\omega-\nu_0)\sin\alpha\cos\gamma],\nonumber\\
\langle\hat{H}_s^{\dagger}\rangle_\theta&=&N_{\theta}^{(r)}\{\lambda_{+}+m^{2}\lambda_{-}+m\sin\alpha(\lambda_+e^{i\gamma}+\lambda_-e^{-i\gamma})\nonumber\\
&=&N_{\theta}^{(r)}[(r\cos\omega+\sqrt{\nu_0})+m^2(r\cos\omega-\sqrt\nu_0)+\nonumber\\
&&2m\sin\alpha(r\cos\omega\cos\gamma+i\sqrt{\nu_0}\sin\gamma)]
\end{eqnarray}
According to Eq. (\ref{A3}), we have
\begin{eqnarray}
\mathcal{F}_{\theta}^{(r)}=\frac{16m^2(s^2-r^2\sin^2{\omega})^2}{[s(1+m^2)+2mr\sin\omega\cos(2\sqrt{\nu_0}\theta+\varphi)]^2}.\label{C1}\nonumber\\
\end{eqnarray}

\begin{figure}[b]
\centering
\includegraphics[width=0.47\textwidth]{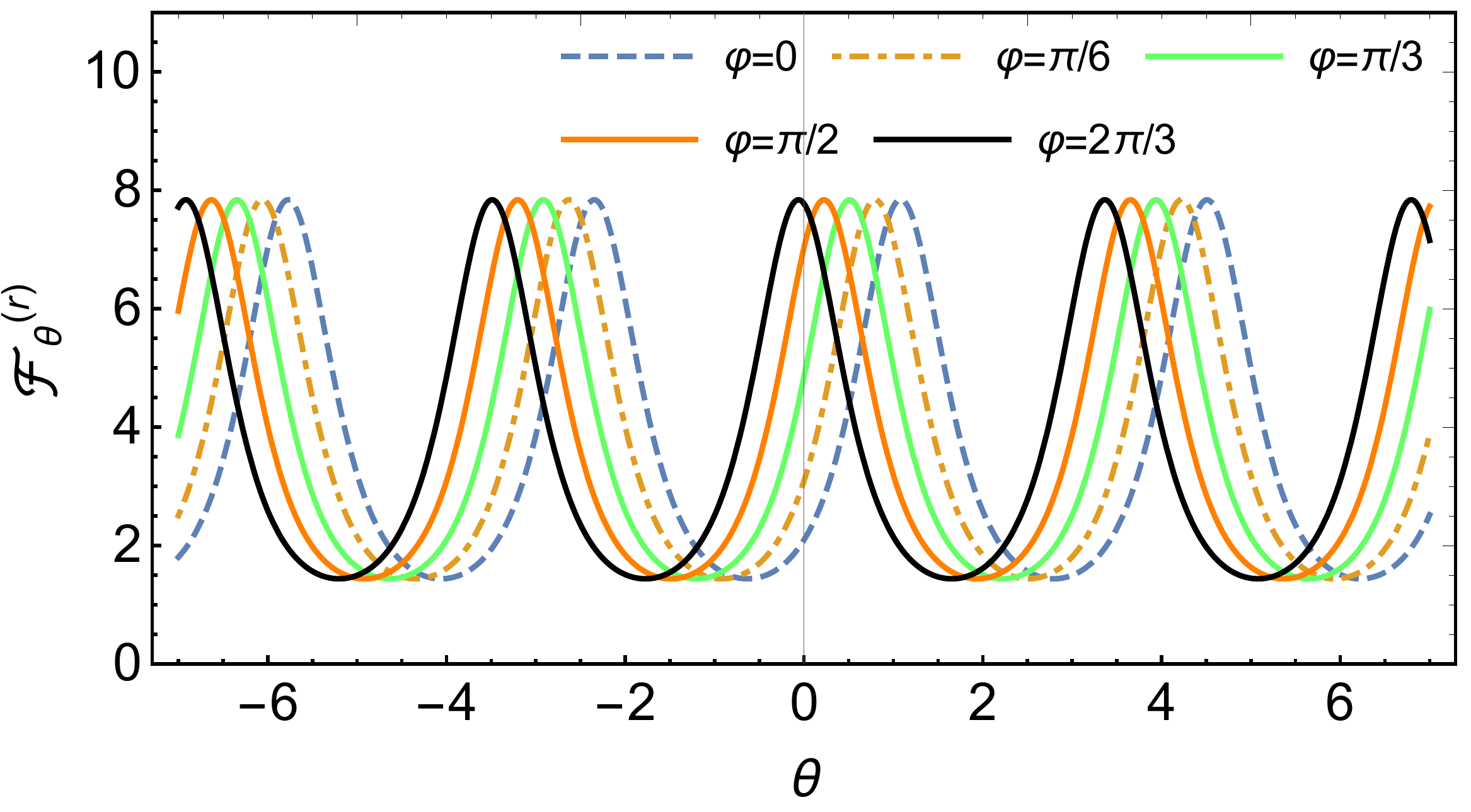}
\caption{The evolution of QFI $\mathcal{F}_{\theta}^{(r)}$ for $\mathcal{PT}$-symmetric Hamiltonian as a function of $\theta$. We set that $s=1$, $r=0.4$, $\omega=\pi/2$, $m=1$ and $\varphi$ changes from $0$ to $2\pi/3$, where $\varphi=$($2\pi/3, \pi/2, \pi/3, \pi/6, 0$) (curves from left to right).}\label{Cfig2}
\end{figure}

In the case of $\mathcal{PT}$-symmetry broken, the eigenvalues turn to $\varepsilon_\pm=\mu\pm\sqrt{\nu_1}=r\cos{\omega}\pm i\sqrt{r^2\sin^2{\omega}-s^2}$, and the normalized eigenstates are
\begin{eqnarray}
\ket{\varepsilon_{\pm}}&=&\frac{1}{\sqrt{(r\sin\omega\pm\sqrt{\nu_1})^2+t^2}}\begin{pmatrix} i(r\sin\omega\pm\sqrt{\nu_1}) \\ t \end{pmatrix}.\nonumber
\end{eqnarray}
Let $\kappa=r\sin\omega/s$, for an arbitrary initial state $\ket{\psi_0}=N(\ket{\varepsilon_+}+me^{i\varphi}\ket{\varepsilon_-})$, similarly, we can obtain
\begin{eqnarray}
N_{\theta}^{(i)}
&=&\frac{|N|^{2}}{K_{\theta}^{(i)}}=\frac{1}{e^{2\sqrt{\nu_1}\theta}+m^{2}e^{-2\sqrt{\nu_1}\theta}+2m\xi\cos\varphi},\nonumber\\
\end{eqnarray}
\begin{eqnarray}
&&\langle \hat{H}_s^{\dagger}\hat{H}_s\rangle_\theta\nonumber\\
&=&N_{\theta}^{(i)}\{|\varepsilon_{+}|^{2}e^{i(\varepsilon_{+}^{\ast}-\varepsilon_{+})\theta}+m^{2}|\varepsilon_{-}|^2e^{i(\varepsilon_{-}^{\ast}-\varepsilon_{-})\theta}\nonumber\\
&&+2\textbf{Re}(me^{i[(\varepsilon_{+}^{\ast}-\varepsilon_{+})\theta+\varphi]}\varepsilon_{+}^{\ast}\varepsilon_{-}\xi)\}\nonumber\\
&=&N_{\theta}^{(i)}\{(r^2-s^2)e^{2\sqrt{\nu_1}\theta}+m^2(r^2-s^2)e^{-2\sqrt{\nu_1}\theta}\nonumber\\
&&+2m\xi[(r^2\cos2\omega+s^2)\cos\varphi+2r\sqrt{\nu_1}\cos\omega\sin\varphi]\},\nonumber\\
\\
&&\langle \hat{H}_s^{\dagger}\rangle_\theta\nonumber\\
&=&N_{\theta}^{(i)}[\varepsilon_{+}^{\ast}e^{2\sqrt{\nu_1}\theta}+m^{2}\varepsilon_{-}^{\ast}e^{-2\sqrt{\nu_1}\theta}+2m\xi\textbf{Re}(\varepsilon_+^*e^{i\varphi})]\nonumber\\
&=&N_{\theta}^{(i)}[(r\cos\omega-i\sqrt{\nu_1})e^{2\sqrt{\nu_1}\theta}\nonumber\\
&&+m^2(r\cos\omega+i\sqrt{\nu_1})e^{-2\sqrt{\nu_1}\theta}\nonumber\\
&&+2m\xi(r\cos\omega\cos\varphi+\sqrt{\nu_1}\sin\varphi)],
\end{eqnarray}
where $\xi=\langle\varepsilon_+|\varepsilon_-\rangle=1/|\kappa|$. Collecting these results, we arrive at
\begin{eqnarray}
\mathcal{F}_{\theta}^{(i)}=\frac{16m^2(r^2\sin^2{\omega}-s^2)^2e^{4\sqrt{\nu_1}\theta}}{[A(e^{4\sqrt{\nu_1}\theta}+m^2)+2mse^{2\sqrt{\nu_1}\theta}\cos\varphi]^2},
\end{eqnarray}
where $A=|r\sin\omega|$. And Fig. \ref{Cfig2} shows that, for the given Hamiltonian, the curve of QFI shifts along the $\theta-$axis with the change of $\varphi$. As discussed in section \uppercase\expandafter{\romannumeral5}, we know that QFI of $\mathcal{PT}$-symmetric Hamiltonian $\hat{H}_s$ can be much larger than that of Hermitian Hamiltonian in some specific intervals of $\theta$, what if the actual parameter $\theta_0$ is in the intervals that QFI of $\hat{H}_s$ is smaller than that of Hermitian Hamiltonians? According to Eq. (\ref{C1}), the relative phase $\varphi$ is only contained in $\cos(2\sqrt{\nu_0}\theta+\varphi)$, hence $\varphi$ has nothing to do with the the amplitude and frequency of $\mathcal{F}_{\theta}^{(r)}$. As shown in Fig. \ref{Cfig2}, we can shift the curve of QFI to maximize it for a specific parameter $\theta_0$ by adjusting the relative phase $\varphi$.

Then we discuss the channel QFI and why it does not reduce to zero as EP is approached. In the two cases of $\mathcal{PT}-$symmetry broken and not, the QFI with optimal initial states respectively are
\begin{eqnarray}
\mathcal{F}_{\theta}^{(r)}&=&\frac{4(s^2-r^2\sin^2{\omega})^2}{[s+r\sin\omega\cos(2\sqrt{\nu_0}\theta+\varphi)]^2},\label{Cmax1}\\
\mathcal{F}_{\theta}^{(i)}&=&\frac{16m^2(r^2\sin^2{\omega}-s^2)^2}{[A(e^{2\sqrt{\nu_1}\theta}+m^2e^{-2\sqrt{\nu_1}\theta})+2ms)]^2},\label{Cmax2}
\end{eqnarray}
where $A=|r\sin\omega|$. For $\mathcal{F}_{\theta}^{(r)}$, if $r\sin\omega>0$, it is maximum when $\cos(2\sqrt{\nu_0}\theta+\varphi)=-1$, then we have
\begin{eqnarray}
\mathcal{F}_{\theta,max}^{(r)}&=&\frac{4(s+r\sin\omega)^2(s-r\sin\omega)^2}{(s-r\sin\omega)^2}\nonumber\\
&=&4(s+r\sin\omega)^2.\label{Cmr1}
\end{eqnarray}
The system reaches EP when $s-r\sin\omega=0$. However, $(s-r\sin\omega)^2$ in the $\mathcal{F}_{\theta,max}^{(r)}$ is omitted, so $\mathcal{F}_{\theta,max}^{(r)}$ will not reduce to zero as EP is approached. If $r\sin\omega<0$, $\mathcal{F}_{\theta}^{(r)}$ is maximum when $\cos(2\sqrt{\nu_0}\theta+\varphi)=1$, we have
\begin{eqnarray}
\mathcal{F}_{\theta,max}^{(r)}&=&\frac{4(s+r\sin\omega)^2(s-r\sin\omega)^2}{(s+r\sin\omega)^2}\nonumber\\
&=&4(s-r\sin\omega)^2.\label{Cmr2}
\end{eqnarray}
Similarly, the system reaches EP when $s+r\sin\omega=0$, but $(s+r\sin\omega)^2$ in $\mathcal{F}_{\theta,max}^{(r)}$ is omitted. To sum up, $\mathcal{F}_{\theta,max}^{(r)}$ can be written as
\begin{eqnarray}
\mathcal{F}_{\theta,max}^{(r)}=4(s+|r\sin\omega|)^2, \label{Cmr}
\end{eqnarray}
it will not reduce to zero as EP is approached.

As for $\mathcal{F}_{\theta}^{(i)}$, it is maximum when $\theta=\ln(-m)/2\sqrt{\nu_1}$, then we have
\begin{eqnarray}
\mathcal{F}_{\theta,max}^{(i)}&=&\frac{4(r^2\sin^2{\omega}-s^2)^2}{(s-|r\sin\omega|)^2}.\nonumber\\
&=&\frac{4(r^2\sin^2{\omega}-s^2)^2B^2}{(r^2\sin^2{\omega}-s^2)^2}=4B^2, \label{Cmi}
\end{eqnarray}
where $B=s+|r\sin\omega|$. Similarly, the system reaches EP when $r^2\sin^2{\omega}-s^2=0$, but $(r^2\sin^2{\omega}-s^2)^2$ is omitted in $\mathcal{F}_{\theta,max}^{(i)}$, so $\mathcal{F}_{\theta,max}^{(i)}$ will not reduce to zero near EP.

\setcounter{equation}{0}
\renewcommand{\theequation}{D\arabic{equation}}
\section*{APPENDIX D: Optimal measurement and sensor of the specific $\mathcal{PT}$-symmetric Hamiltonian}

In the main text, we give an optimal measurement $|0\rangle\langle0|$ for the $\mathcal{PT}$-symmetric non-Hermitian Hamiltonian $\hat{H}_s$ in the case of PT symmetry not broken, in this section, we are going to prove that the measurement $|0\rangle\langle0|$ satisfies the condition Eq. (\ref{B4}).  Here, we denote $\hat{X}=|0\rangle\langle0|$ and use $\kappa=r/s$, and remember that we set $\omega=\pi/2$. As discussed in the main text, the measurement is optimal only for the optimal initial state. In the case of $\mathcal{PT}$-symmetry not broken, the optimal initial state is $\ket{\psi_{0}}=N(\ket{\lambda_{+}}+e^{i\varphi}\ket{\lambda_{-}})$, then we can obtain $\langle\hat{H_s}\rangle_\theta$ and $\langle\hat{X}\rangle_\theta$ as follows:
\begin{eqnarray}
\langle\hat{H_s}\rangle_\theta&=&-i2N_\theta^{(r)}\nu_0[\kappa\sqrt{1-\kappa^2}\cos(2\sqrt{\nu_0}\theta+\varphi)+\nonumber\\
&&\kappa^2\sin(2\sqrt{\nu_0}\theta+\varphi)],\nonumber\\
\langle \hat{X}\rangle_\theta&=&N_\theta^{(r)}\{1+[(2\kappa^2-1)\cos(2\sqrt{\nu_0}\theta+\varphi)-\nonumber\\
&&2\kappa\sqrt{1-\kappa^2}\sin(2\sqrt{\nu_0}\theta+\varphi)]\},
\end{eqnarray}
According to Eq. (\ref{B5}) and Eq. (\ref{B6}), we have
\begin{eqnarray}
\ket{f}&=&(\hat{H_s}-\langle \hat{H_s}\rangle)\ket{\varphi_\theta}\nonumber\\
&=&\frac{4e^{i\frac{\varphi}{2}}N_\theta^{(r)}\nu_0}{s}\sqrt{\frac{N_\theta^{(r)}}{2}}\begin{pmatrix} \cos\gamma_0 \\ i\sin(\gamma_0+\beta) \end{pmatrix},\\
\ket{g}&=&(\hat{X}-\langle \hat{X}\rangle)\ket{\varphi_\theta}\nonumber\\
&=&4e^{i\frac{\varphi}{2}}N_\theta^{(r)}\sqrt{\frac{N_\theta^{(r)}}{2}}\begin{pmatrix} i\cos^2\gamma_0\sin(\gamma_0+\beta) \\ -\cos\gamma_0\sin^2(\gamma_0+\beta) \end{pmatrix},\nonumber\\
\end{eqnarray}
where $\gamma_0=\sqrt{\nu_0}\theta+\varphi/2$ and $\beta$ is determined by $\kappa$ ($\sin\beta=\kappa, \cos\beta=-\sqrt{1-\kappa^{2}}$). Then we obtain
\begin{eqnarray}
\ket{f}=-\frac{i\nu_0}{s\cos\gamma_0\sin(\gamma_0+\beta)}\ket{g}.
\end{eqnarray}
Obviously, this result satisfies the condition $\ket{f}=iC\ket{g}$. Therefore, $|0\rangle\langle0|$ is indeed the optimal measurement only for the optimal initial state.

\begin{figure}[t]
\centering
\includegraphics[ width=0.455\textwidth]{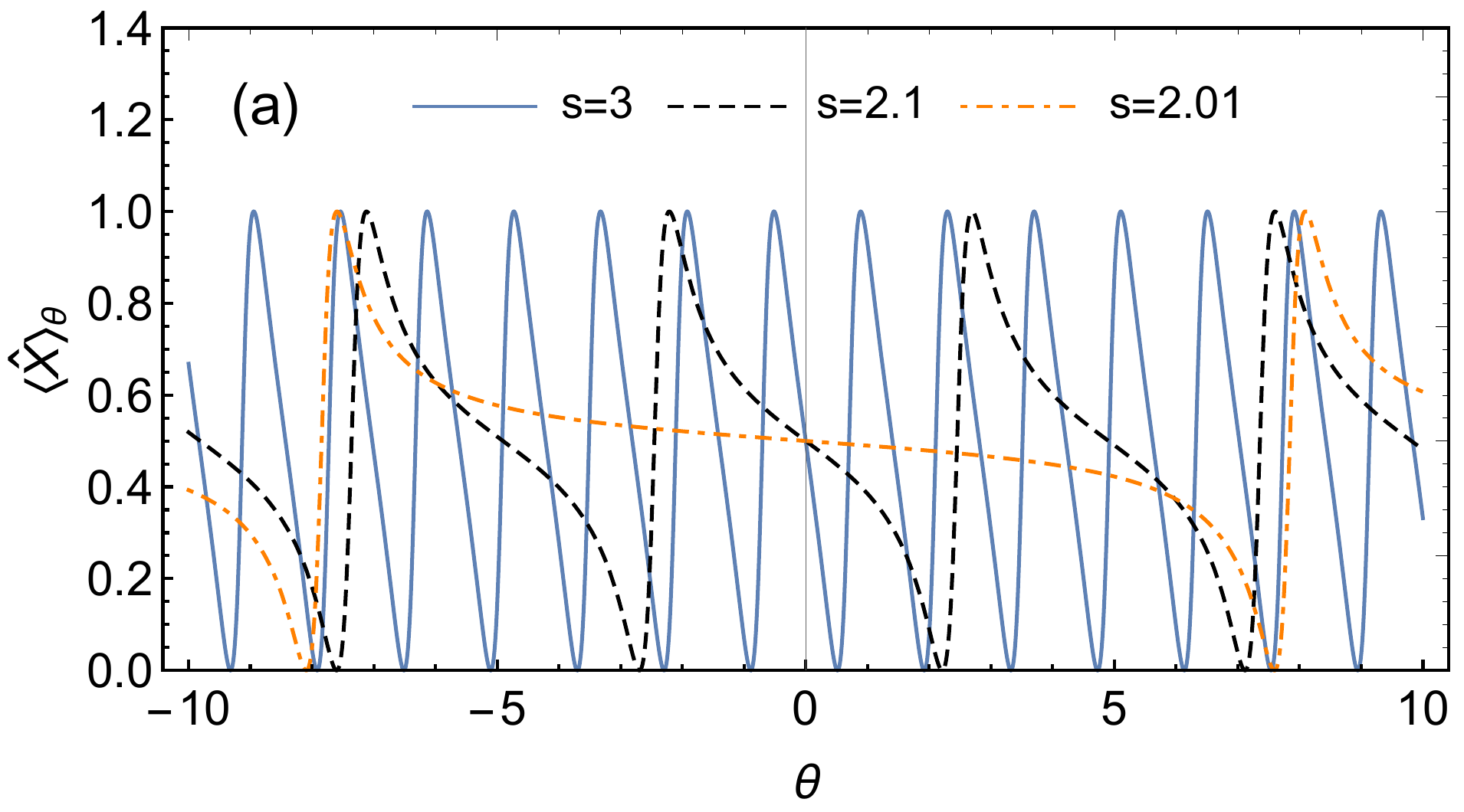}
\includegraphics[ width=0.455\textwidth]{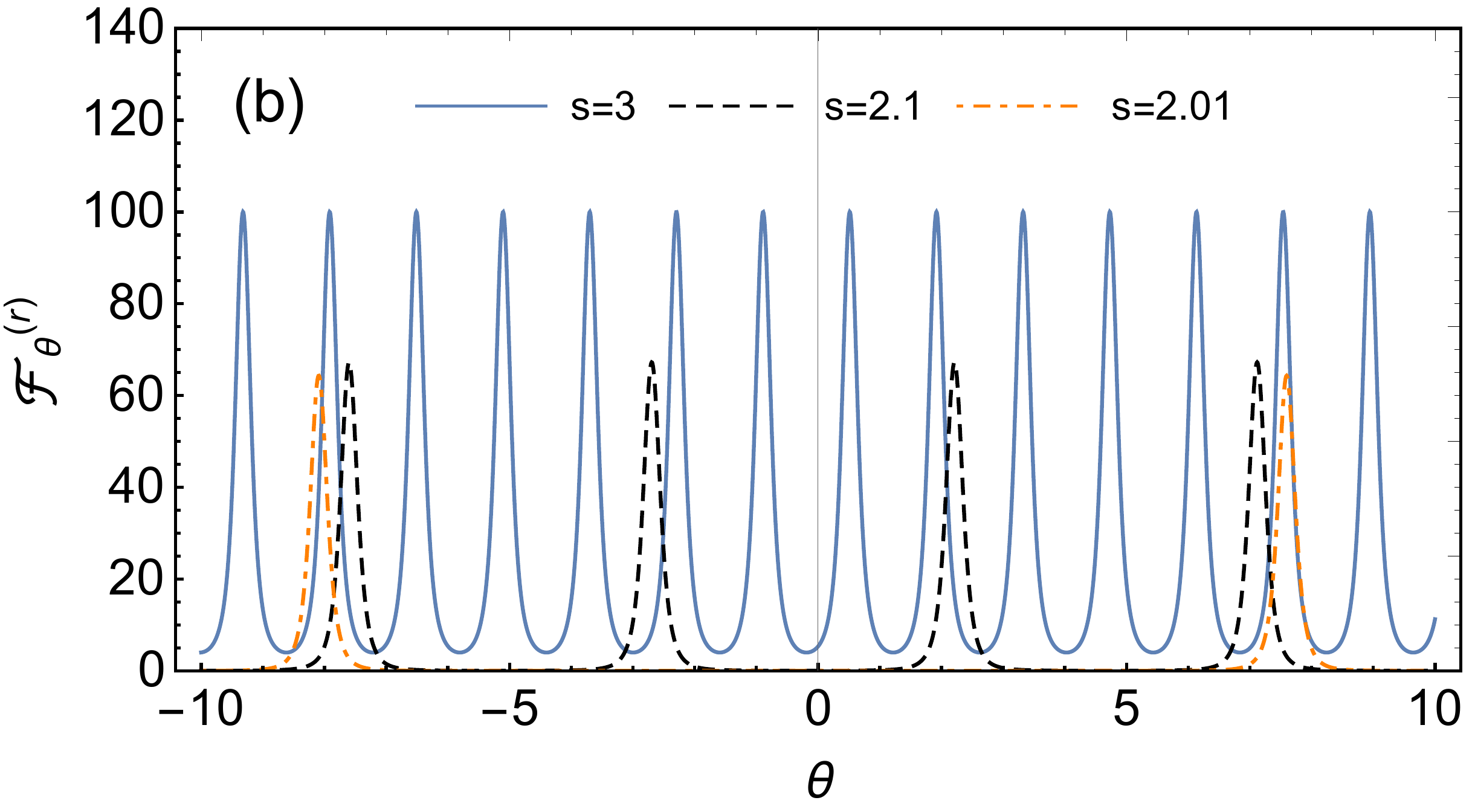}
\caption{Quantum sensor of $\mathcal{PT}$-symmetric non-Hermitian Hamiltonian. (a) The expectation value $\langle\hat{X}\rangle$ as functions of evolution parameter $\theta$, we change the value of $s$, so that the system gradually approaches EP. We set $r=2,\omega=\pi/2$ and the initial state is $\ket{\psi_0}=N(\ket{\lambda_+}+\ket{\lambda_-})$, the system reaches EP when $s=2$, where orange dot-dashed line ($s=2.01$), black dashed line ($s=2.1$) and blue solid line ($s=3$). (b) The QFI as functions of evolution parameter $\theta$, the parameter setting is the same as (a).}\label{fig6}
\end{figure}

As discussed in the main text, the QFI $\mathcal{F}_{\theta}$ is always zero at EP if the initial states can be expressed with eigenstates, so we can speculate that the sensitivity of sensors reduces as EP is approached. We use the optimal measurement $\hat{X}=|0\rangle\langle0|$ as a sensor to verify it, keep the initial state optimal, i.e., $\ket{\psi_0}=N(\ket{\lambda_+}+\ket{\lambda_-})$, the expectation of $\hat{X}$ is
\begin{eqnarray}
\langle\hat{X}\rangle_\theta=\frac{1-\sin(2\theta\sqrt{s^2-r^2\sin^2\omega}-\alpha)}{2+2\sin\alpha\cos(2\theta\sqrt{s^2-r^2\sin^2\omega})}.
\end{eqnarray}
In Fig. \ref{fig6}(a), we show the expectation values $\langle\hat{X}\rangle_\theta$ and the QFI with different values of $s$. We can see that $\langle\hat{X}\rangle$ is less sensitive to the changes in parameter $\theta$, as the EP is approached, i.e., the sensitivity of sensor $\hat{X}$ reduces near EP. Note that in Fig. \ref{fig6}(a), there are still some intervals that $\hat{X}$ is sensitive to the parameter $\theta$. As we discussed above, the channel QFI $\mathcal{F}_{\theta,max}$ will not reduce to zero near EP, it corresponds to the intervals that $\hat{X}$ is sensitive to parameter $\theta$. Comparing Fig. \ref{fig6}(a) with Fig. \ref{fig6}(b), we can see that the intervals with high sensor sensitivity corresponds to the intervals with large QFI. The performance of sensor $\hat{X}$ is consistent with our analysis of QFI.\\
\begin{figure}[t]
\centering
\includegraphics[ width=0.455\textwidth]{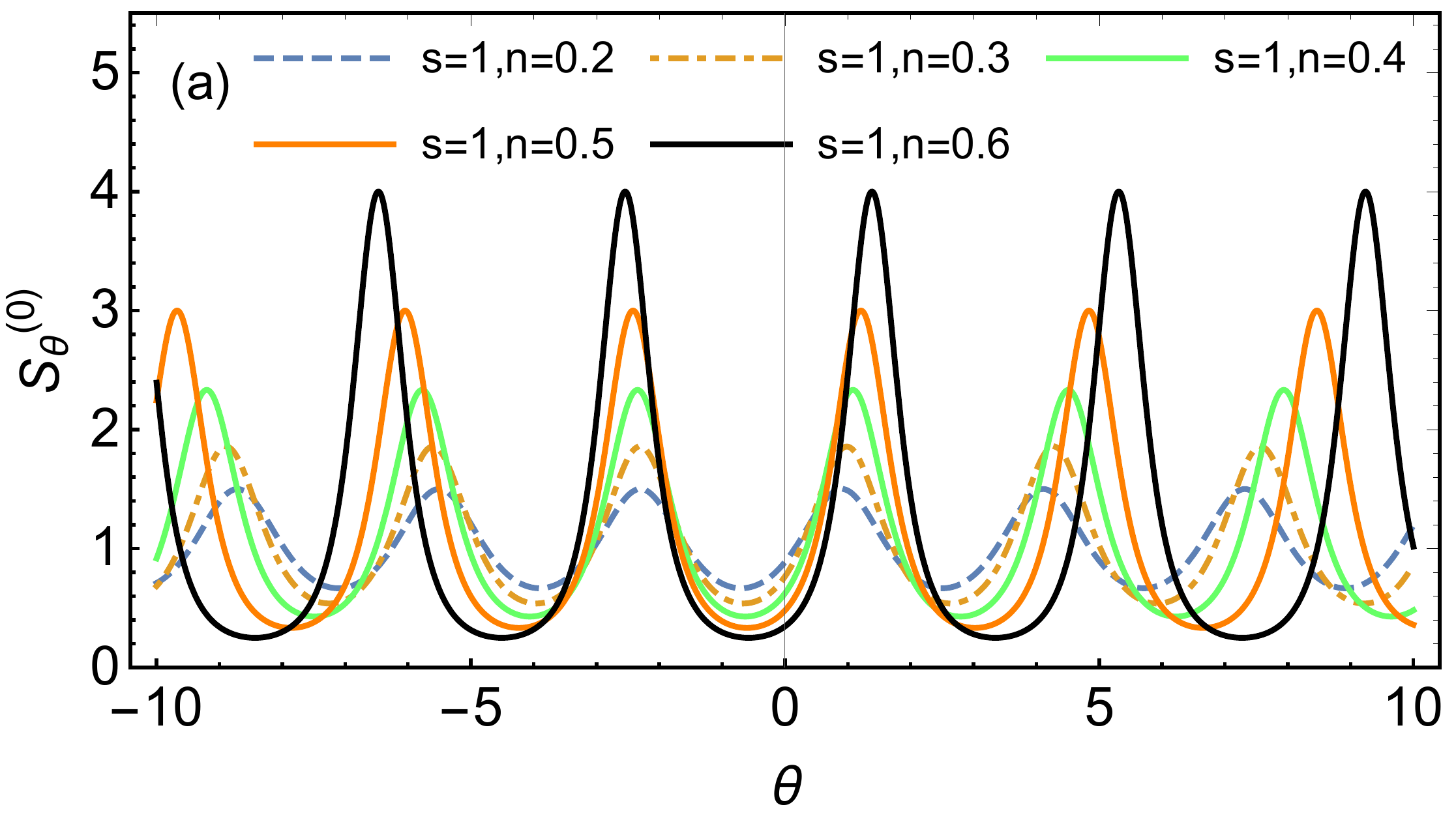}
\includegraphics[ width=0.47\textwidth]{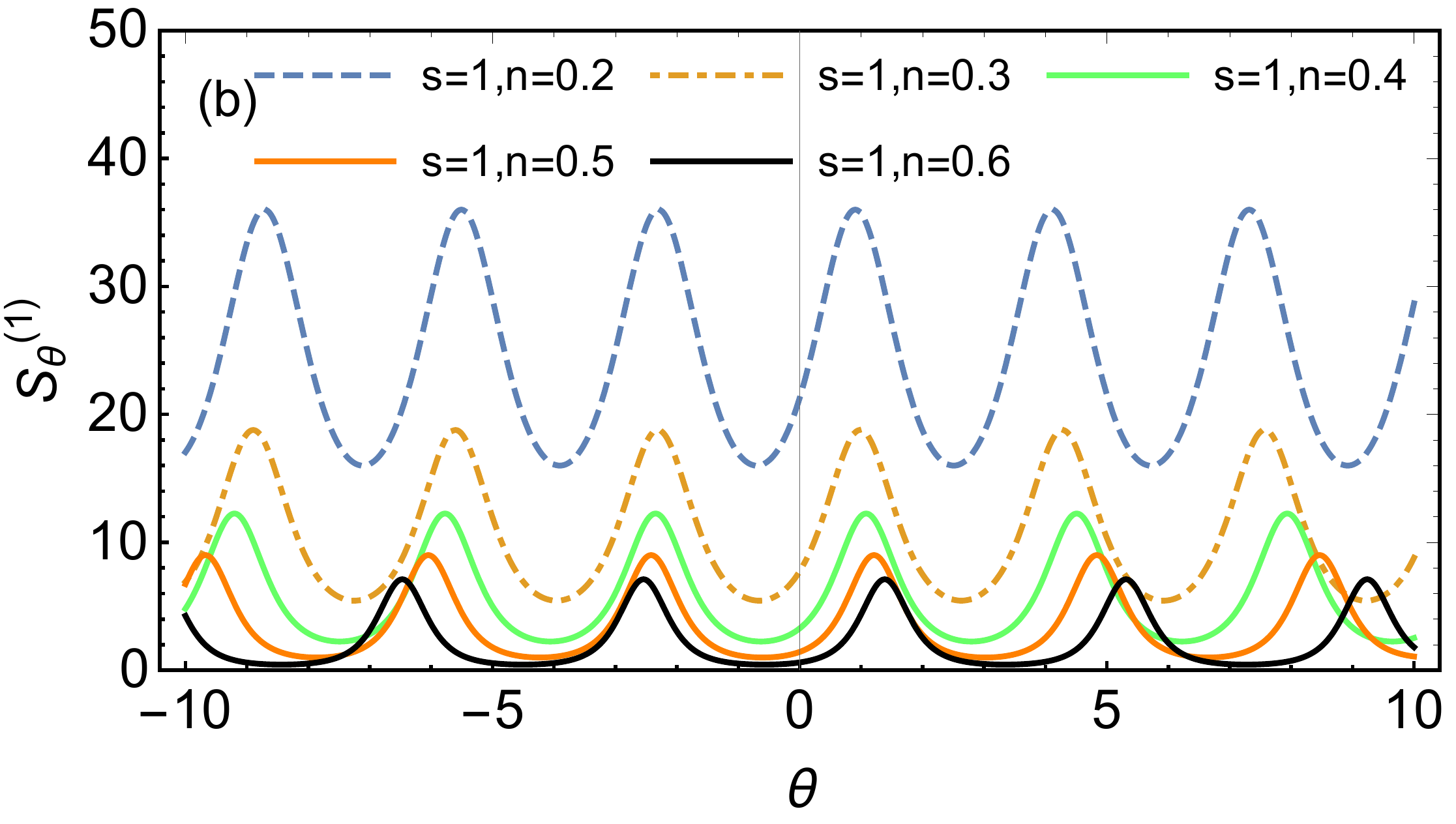}
\caption{The evolution of ratio $S_{\theta}$ as a function of $\theta$. We set that $s=1$, $\kappa$ changes from $0.2$ to $0.6$ with an interval of $0.2$, and the initial state $\ket{\psi_{0}}=N(\ket{\lambda_{+}}+\ket{\lambda_{-}})$ remains unchanged. (a) The ratio factor $S_{\theta}^{(0)}$ of systems with different modulus of energy, where $n=$ ($0.2, 0.3, 0.4, 0.5, 0.6$) (from bottom to top). (b) The ratio factor $S_{\theta}^{(1)}$ of systems with different modulus of energy, where $n=$ ($0.2, 0.3, 0.4, 0.5, 0.6$) (curves from top to bottom).}\label{fig7}
\end{figure}

\setcounter{equation}{0}
\renewcommand{\theequation}{E\arabic{equation}}
\section*{APPENDIX E: Comparison of Hermitian and non-Hermitian Hamiltonian QFI}
Knowing the QFI of $\hat{H}_s$, we compare $\mathcal{F}_{\theta}^{(r)}$ with the QFI of two Hermitian Hamiltonians, for simplicity, we set $\omega=\pi/2$. The Hermitian Hamiltonians are
\begin{eqnarray}
\hat{H}_0=\begin{pmatrix} \sqrt{s^2-r^2} & 0 \\ 0 & -\sqrt{s^2-r^2} \end{pmatrix}, \hat{H}_1=\begin{pmatrix} s & -ir \\ ir & s \end{pmatrix}.\nonumber
\end{eqnarray}
Note that $\hat{H}_0$ has the same eigenvalues with $\mathcal{PT}$-symmetric Hamiltonian $\hat{H}_s$, and $\hat{H}_1^{\dagger}\hat{H}_1=\hat{H}_s^{\dagger}\hat{H}_s$. That is why we select these two Hermitian Hamiltonians.
It is not difficult to calculate the corresponding optimal QFI, we have $\mathcal{F}_{\theta}^{(0)}=4\lambda_0^2$ and $\mathcal{F}_{\theta}^{(1)}=4r^2$. We find that the QFI of Hamiltonian $\hat{H}_s$ can be much larger than that of two Hermitian Hamiltonians. To illustrate this, we define the ratios $S_\theta$ as functions as follows:
\begin{eqnarray}
S_{\theta}^{(0)}&=&\frac{\mathcal{F}_{\theta}^{(r)}}{\mathcal{F}_{\theta}^{(0)}}=\frac{4m^2(1-\kappa^2)}{[1+m^2+2m\kappa\sin(2\sqrt{\nu_0}\theta+\varphi+\alpha)]^2},\nonumber\\
S_{\theta}^{(1)}&=&\frac{\mathcal{F}_{\theta}^{(r)}}{\mathcal{F}_{\theta}^{(1)}}=\frac{4m^2(1-\kappa^2)^2}{\kappa^2[1+m^2+2m\kappa\sin(2\sqrt{\nu_0}\theta+\varphi+\alpha)]^2}.\nonumber\\
\end{eqnarray}
Let us consider an example of the comparison, we set the initial state as $\ket{\psi_{0}}=N(\ket{\lambda_{+}}+\ket{\lambda_{-}})$ which is optimal, and change the modulus of energy $\lambda_{0}$. If there is $S_{\theta}>1$, our conjecture is verified. We can see that in Fig. \ref{fig7}(a) and (b), the ratios $S_{\theta}^{(0)}$ and $S_{\theta}^{(1)}$ change with the value of $n$, and the curves above the grey dotted line indicate that the QFI for non-Hermitian system can be larger than that of Hermitian systems in some intervals, among them, the maximum value of $S_{\theta}^{(0)}$ can reach 4, while the maximum value of $S_{\theta}^{(1)}$ can even reach 36! As mentioned in section \uppercase\expandafter{\romannumeral3}, as we change the relative phase $\varphi$ in initial state $\ket{\psi_0}$, the curves shift along the $\theta-$axis, so that we can guarantee that $\mathcal{F}_{\theta}^{(r)}$ can be larger than $\mathcal{F}_{\theta}^{(0)}$ or $\mathcal{F}_{\theta}^{(1)}$ in arbitrary intervals of $\theta$.

\setcounter{equation}{0}
\renewcommand{\theequation}{F\arabic{equation}}
\section*{APPENDIX F: QFI of a two coupled bosonic modes system}

In most cases, the dynamic of an open system is described with master equation, however, if the quantum jump terms can be neglect, the system can be described with an effective non-Hermitian Hamiltonian $H_{eff}$ \cite{qtj}. Our results is still feasible for such a system.

For example, let us consider a two coupled bosonic modes system
\begin{eqnarray}
\hat{H}=\omega(\hat{a}^\dagger\hat{a}+\hat{b}^\dagger \hat{b})+g(\hat{a}^\dagger \hat{b}+\hat{b}^\dagger \hat{a}),
\end{eqnarray}
and the dynamic is
\begin{eqnarray}
\frac{\partial\hat{\rho}(t)}{\partial t}=-i[\hat{H},\hat{\rho}(t)]+(\frac{\gamma_a}{2}\mathcal{D}[\hat{a}]+\frac{\gamma_b}{2}\mathcal{D}[\hat{b}])\hat{\rho}(t),\nonumber\\
\end{eqnarray}
where the dissipator $\mathcal{D}[\hat{a}]\hat{\rho}(t)=\hat{a}\hat{\rho}(t)\hat{a}^\dagger-\{\hat{a}^\dagger\hat{a},\hat{\rho}(t)\}/2$. In the semiclassical limit, this system can be described with a non-Hermitian Hamiltonian as follow \cite{qtj}
\begin{eqnarray}
\hat{H}_{eff}=(\omega-\frac{i\gamma_a}{2})\hat{a}^\dagger \hat{a}+(\omega-\frac{i\gamma_b}{2})\hat{b}^\dagger \hat{b}+g(\hat{a}^\dagger \hat{b}+\hat{b}^\dagger \hat{a}).\label{ff1}\nonumber\\
\end{eqnarray}
In the case of one excitation, the eigenvalues of the system are $\lambda_\pm=\omega-i\overline{\gamma}/2\pm\xi$, where $\overline{\gamma}=(\gamma_a+\gamma_b)/2$, $\xi^2=g^2-\gamma^2/4$ and $\gamma=(\gamma_a-\gamma_b)$. Then the normalized eigenstates can be expressed with Fock state as follow
\begin{eqnarray}
|\lambda_\pm\rangle=\frac{1}{\sqrt{2}g}[(-i\frac{\gamma}{2}\pm\xi)|0,1\rangle+g|1,0\rangle].
\end{eqnarray}
The dynamic of this non-Hermitian Hamiltonian is $d\hat{\rho}_\theta/d\theta=-i(\hat{H}_{eff}\hat{\rho}_\theta-\hat{\rho}_\theta \hat{H}_{eff}^\dagger)$, then the evolution operator is $\hat{U}_\theta=e^{-i\hat{H}\theta}$. Here, we assume that $\xi^2=g^2-\gamma^2/4\geq0$, for a given normalized initial state $|\psi_0\rangle=(|\lambda_{+}\rangle-|\lambda_{-}\rangle)g/(\sqrt{2}\xi)=|0,1\rangle$, the QFI is
\begin{eqnarray}
\mathcal{F}_\theta&=&4(\langle\varphi_\theta|H^\dagger H|\varphi_\theta\rangle-\langle\varphi_\theta|H^\dagger|\varphi_\theta\rangle\langle\varphi_\theta|H|\varphi_\theta\rangle) \nonumber\\
&=&\frac{64g^2\xi^4}{[-4g^2+\gamma^2\cos(2\xi\theta)+2\gamma\xi\sin(2\xi\theta)]^2}.
\end{eqnarray}
Therefore, our result can also be used for open systems which can be described with an effective non-Hermitian Hamiltonian in the semiclassical limit.


\begin{thebibliography}{99}

\bibitem{rev1} C. W. Helstrom, \textit{Quantum Detection and Estimation Theory} (Academic Press, New Y ork, 1976).

\bibitem{rev2} A. S. Holevo, \textit{Probabilistic and Statistical Aspects of Quantum Theory} (North-Holland, Amsterdam, 1982).

\bibitem{rev3} S. L. Braunstein and C. M. Caves, \href{https://doi.org/10.1103/PhysRevLett.72.3439}{Phys. Rev. Lett. \textbf{72}, 3439 (1994)}.

\bibitem{rev4} S. L. Braunstein, C. M. Caves, G.J. Milburn, \href{https://doi.org/10.1006/aphy.1996.0040.}{Ann. Phys. \textbf{247}, 135-173 (1996)}.


\bibitem{rev5} L. Pezz\'{e} and A. Smerzi, \href{https://doi.org/10.1103/PhysRevLett.102.100401}{Phys. Rev. Lett. \textbf{102}, 100401 (2009)}.

\bibitem{rev6} P. Hyllus, W. Laskowski, R. Krischek, C. Schwemmer, W. Wieczorek, H. Weinfurter, L. Pezz\'{e}, and A. Smerzi, \href{https://doi.org/10.1103/PhysRevA.85.022321}{Phys. Rev. A \textbf{85}, 022321 (2012)}.

\bibitem{rev7} G. T\'{o}th, \href{https://doi.org/10.1103/PhysRevA.85.022322}{Phys. Rev. A \textbf{85}, 022322 (2012)}.

\bibitem{rev8} A. del Campo, I. L. Egusquiza, M. B. Plenio, and S. F. Huelga, \href{https://doi.org/10.1103/PhysRevLett.110.050403}{Phys. Rev. Lett. \textbf{110}, 050403 (2013)}.

\bibitem{rev9} M. M. Taddei, B. M. Escher, L. Davidovich, and R. L. de Matos Filho, \href{https://doi.org/10.1103/PhysRevLett.110.050402}{Phys. Rev. Lett. \textbf{110}, 050402 (2013)}.

\bibitem{rev10} F. Fr\"owis, \href{https://doi.org/10.1103/PhysRevA.85.052127}{Phys. Rev. A \textbf{85}, 052127 (2012)}.

\bibitem{rev11} F. Fr\"owis and W. D\"ur, \href{https://doi.org/10.1088/1367-2630/14/9/093039}{New J. Phys. \textbf{14} 093039 (2012)}.

\bibitem{rev12} F. Fr\"owis and W. D\"ur, \href{https://doi.org/10.1103/PhysRevLett.106.110402}{Phys. Rev. Lett. \textbf{106}, 110402 (2011)}.

\bibitem{rev13} A. Smerzi, \href{https://doi.org/10.1103/PhysRevLett.109.150410}{Phys. Rev. Lett. \textbf{109}, 150410 (2012)}.

\bibitem{rev14} F. Sch\"{a}fer, I. Herrera, S. Cherukattil, C. Lovecchio, F.S. Cataliotti, F. Caruso and A. Smerzi, \href{https://doi.org/10.1038/ncomms4194}{Nat. Commun\textbf{5}, 3194 (2014)}.

\bibitem{rev15} C. M. Bender and S. Boettcher, \href{https://doi.org/10.1103/PhysRevLett.80.5243}{Phys. Rev. Lett. \textbf{80}, 5243 (1998)}.

\bibitem{rev16} P. Dorey, C. Dunning and R. Tateo, \href{https://doi.org/10.1088/0305-4470/34/28/305}{J. Phys. A: Math. Gen. \textbf{34} 5679 (2001)}. 

\bibitem{rev17} C. M. Bender, D. C. Brody, and H. F. Jones, \href{https://doi.org/10.1103/PhysRevLett.89.270401}{Phys. Rev. Lett. \textbf{89}, 270401 (2002)}.

\bibitem{rev18} A. Mostafazadeh, \href{https://doi.org/10.1063/1.1461427}{J. Math. Phys. \textbf{43}, 2814 (2002)}.

\bibitem{rev19} C. M. Bender, \href{https://doi.org/10.1088/0034-4885/70/6/R03}{Rep. Prog. Phys. \textbf{70} 947 (2007)}.

\bibitem{rev20} M.V. Berry, \href{https://doi.org/10.1023/B:CJOP.0000044002.05657.04}{Czech. J. Phys. \textbf{54}, 1039-1047 (2004)}.

\bibitem{rev21} W. D. Heiss, \href{https://doi.org/10.1088/1751-8113/45/44/444016}{J. Phys. A: Math. Theor. \textbf{45} 444016 (2012)}.

\bibitem{rev22} N. Lazarides and G. P. Tsironis, \href{https://doi.org/10.1103/PhysRevLett.110.053901}{Phys. Rev. Lett. \textbf{110}, 053901 (2013)}.

\bibitem{rev23} F. Monticone, C. A. Valagiannopoulos, and A. Al\`{u}, \href{https://doi.org/10.1103/PhysRevX.6.041018}{Phys. Rev. X \textbf{6}, 041018 (2016)}.

\bibitem{rev24} Y.-C. Hu and T. L. Hughes, \href{https://doi.org/10.1103/PhysRevB.84.153101}{Phys. Rev. B \textbf{84}, 153101 (2011)}.

\bibitem{rev25} M. S. Rudner and L. S. Levitov, \href{https://doi.org/10.1103/PhysRevLett.102.065703}{Phys. Rev. Lett. \textbf{102}, 065703 (2009)}.

\bibitem{rev26} K. G. Makris, R. El-Ganainy, D. N. Christodoulides, and Z. H. Musslimani, \href{https://doi.org/10.1103/PhysRevLett.100.103904}{Phys. Rev. Lett. \textbf{100}, 103904 (2008)}.

\bibitem{rev27} R. El-Ganainy, K. G. Makris, D. N. Christodoulides, and Z. H. Musslimani, \href{https://doi.org/10.1364/OL.32.002632}{Opt. Lett. \textbf{32}, 2632-2634 (2007)}.

\bibitem{rev28} Z. H. Musslimani, K. G. Makris, R. El-Ganainy, and D. N. Christodoulides, \href{https://doi.org/10.1103/PhysRevLett.100.030402}{Phys. Rev. Lett. \textbf{100}, 030402 (2008)}.

\bibitem{rev29} A. Guo, G. J. Salamo, D. Duchesne, R. Morandotti, M. Volatier-Ravat, V. Aimez, G. A. Siviloglou, and D. N. Christodoulides, \href{https://doi.org/10.1103/PhysRevLett.103.093902}{Phys. Rev. Lett. \textbf{103}, 093902 (2009)}.

\bibitem{rev30} C. E. R\"{u}ter, K. G. Makris, R. El-Ganainy, D. N. Christodoulides, M. Segev and D. Kip, \href{https://doi.org/10.1038/nphys1515}{Nature Phys \textbf{6}, 192-195 (2010)}.

\bibitem{hev1} S. Wildermuth, S. Hofferberth, I. Lesanovsky, S. Groth, P. Kr\"{u}ger, and J. Schmiedmayer, \href{https://doi.org/10.1063/1.2216932}{Appl. Phys. Lett. \textbf{88}, 264103 (2006)}.

\bibitem{hev2} M. Vengalattore, J. M. Higbie, S. R. Leslie, J. Guzman, L. E. Sadler, and D. M. Stamper-Kurn, \href{https://doi.org/10.1103/PhysRevLett.98.200801}{Phys. Rev. Lett. \textbf{98}, 200801 (2007)}.

\bibitem{hev3} N. Behbood, F. Martin Ciurana, G. Colangelo, M. Napolitano, M. W. Mitchell, and R. J. Sewell, \href{https://doi.org/10.1063/1.4803684}{Appl. Phys. Lett. \textbf{102}, 173504 (2013)}.

\bibitem{hev4} M. Koschorreck, M. Napolitano, B. Dubost, and M. W. Mitchell, \href{https://doi.org/10.1063/1.3555459}{Appl. Phys. Lett. \textbf{98}, 074101 (2011)}.

\bibitem{hev5} W. Muessel, H. Strobel, D. Linnemann, D. B. Hume, and M. K. Oberthaler, \href{https://doi.org/10.1103/PhysRevLett.113.103004}{Phys. Rev. Lett. \textbf{113}, 103004 (2014)}.

\bibitem{hev6} W. Wasilewski, K. Jensen, H. Krauter, J. J. Renema, M. V. Balabas, and E. S. Polzik, \href{https://doi.org/10.1103/PhysRevLett.104.103602}{Phys. Rev. Lett. \textbf{104}, 209902(E) (2010)}.

\bibitem{hev7} P. M. Anisimov, G. M. Raterman, A. Chiruvelli, W. N. Plick, S. D. Huver, H. Lee, and J. P. Dowling, \href{https://doi.org/10.1103/PhysRevLett.98.200801}{Phys. Rev. Lett. \textbf{104}, 103602 (2010)}.

\bibitem{hev8} J. Joo, W. J. Munro, and T. P. Spiller, \href{https://doi.org/10.1103/PhysRevLett.107.083601}{Phys. Rev. Lett. \textbf{107}, 219902(E) (2011)}.

\bibitem{hev9} M. G. Genoni, S. Olivares, and M. G. A. Paris, \href{https://doi.org/10.1103/PhysRevLett.106.153603}{Phys. Rev. Lett. \textbf{106}, 153603 (2011)}.

\bibitem{hev10} N. Thomas-Peter, B. J. Smith, A. Datta, L. Zhang, U. Dorner, and I. A. Walmsley, \href{https://doi.org/10.1103/PhysRevLett.107.113603}{Phys. Rev. Lett. 107, 113603 (2011)}.

\bibitem{hev11} L. Pezz\`{e}, M. A. Ciampini, N. Spagnolo, P. C. Humphreys, A. Datta, I. A. Walmsley, M. Barbieri, F. Sciarrino, and A. Smerzi, \href{https://doi.org/10.1103/PhysRevLett.119.130504}{Phys. Rev. Lett. \textbf{119}, 130504 (2017)}.

\bibitem{hev12} X.-M. Lu and X. Wang, \href{https://doi.org/10.1103/PhysRevLett.126.120503}{Phys. Rev. Lett. \textbf{126}, 120503 (2021)}.

\bibitem{hev13} H. Yuan and C.-H. F. Fung, \href{https://doi.org/10.1103/PhysRevLett.115.110401}{Phys. Rev. Lett. \textbf{115}, 110401 (2015)}.

\bibitem{hev14} H. Yuan, \href{https://doi.org/10.1103/PhysRevLett.117.160801}{Phys. Rev. Lett. \textbf{117}, 160801 (2016)}.

\bibitem{rev31} J. Wiersig, \href{https://doi.org/10.1103/PhysRevLett.112.203901}{Phys. Rev. Lett. \textbf{112}, 203901 (2014)}.

\bibitem{rev32} Z.-P. Liu, J. Zhang, \c{S}. K. \"{o}zdemir, B. Peng, H. Jing, X.-Y. L\"{u}, C.-W. Li, L. Yang, F. Nori, and Y.-X. Liu, \href{https://doi.org/10.1103/PhysRevLett.117.110802}{Phys. Rev. Lett. \textbf{117}, 110802 (2016)}.

\bibitem{rev33} W. Chen, \c{S}. K. \"{o}zdemir, G. Zhao, J. Wiersig, L. Yang, \href{https://doi.org/10.1038/nature23281}{Nature \textbf{548}, 192 (2017)}.

\bibitem{rev34} H. Hodaei, A. U. Hassan, S. Wittek, H. Garcia-Gracia, R. El-Ganainy, D. N. Christodoulides, M. Khajavikhan, \href{https://doi.org/10.1038/nature23280}{Nature \textbf{548}, 187 (2017)}.

\bibitem{nev1} S. Yu, Y. Meng, J.-S. Tang, X.-Y. Xu, Y.-T. Wang, P. Yin, Z.-J. Ke, W. Liu, Z.-P. Li, Y.-Z. Yang, G. Chen, Y.-J. Han, C.-F. Li, and G.-C. Guo, \href{https://doi.org/10.1103/PhysRevLett.125.240506}{Phys. Rev. Lett. \textbf{125}, 240506 (2020)}.

\bibitem{nev2} Y. Chu, Y. Liu, H. Liu, and J. Cai, \href{https://doi.org/10.1103/PhysRevLett.124.020501}{Phys. Rev. Lett. \textbf{124}, 020501 (2020)}.

\bibitem{nev3} C. Chen, L. Jin and R.-B. Liu, \href{https://doi.org/10.1088/1367-2630/ab32ab}{New J. Phys. \textbf{21} 083002 (2019)}.

\bibitem{nev4} W. Langbein, \href{https://doi.org/10.1103/PhysRevA.98.023805}{Phys. Rev. A \textbf{98}, 023805 (2018)}.

\bibitem{nev5} L. Bao, B. Qi, D. Dong, and F. Nori, \href{https://doi.org/10.1103/PhysRevA.103.042418}{Phys. Rev. A \textbf{103}, 042418 (2021)}.

\bibitem{nev6} M. Zhang, W. Sweeney, C. W. Hsu, L. Yang, A.D. Stone, and L. Jiang, \href{https://doi.org/10.1103/PhysRevLett.123.180501}{Phys. Rev. Lett. \textbf{123}, 180501  (2019)}.

\bibitem{nev7} J. Wang, D. Mukhopadhyay, and G. S. Agarwal, \href{https://doi.org/10.1103/PhysRevResearch.4.013131}{Phys. Rev. Research \textbf{4}, 013131 (2022)}.

\bibitem{opm1} H.-K. Lau, A. A. Clerk, \href{https://doi.org/10.1038/s41467-018-06477-7}{Nat Commun \textbf{9}, 4320 (2018)}.

\bibitem{emp1} D. Xie, C. Xu and A. M. Wang, \href{https://doi.org/10.1016/j.rinp.2021.104430}{Results Phys \textbf{26}, 104430 (2021)}.

\bibitem{emnh} J. Li, H. Liu, Z. Wang, X. Yi, \href{https://doi.org/10.48550/arXiv.2103.07099}{arXiv:2103.07099}.

\bibitem{cqfi} J. Ko{\l}ody\'{n}ski and R. Demkowicz-Dobrza\'{n}ski, \href{https://doi.org/10.48550/arXiv.2103.07099}{New J. Phys. \textbf{15}
    073043 (2013)}.

\bibitem{qtj} F. Minganti, A. Miranowicz, R. W. Chhajlany, and F. Nori, \href{https://doi.org/10.1103/PhysRevA.100.062131}{Phys. Rev. A \textbf{100}, 062131 (2019)}.

\bibitem{spn1} J.-S. Tang, Y.-T. Wang, S. Yu, D.-Y. He, J.-S. Xu, B.-H. Liu, G. Chen, Y.-N. Sun, K. Sun, Y.-J. Han, C.-F. Li and G.-C. Guo, \href{https://doi.org/10.1038/nphoton.2016.144}{Nature Photon \textbf{10}, 642 (2016)}.

\bibitem{spn2} Q. Li, C.-J. Zhang, Z.-D. Cheng, W.-Z. Liu, J.-F. Wang, F.-F. Yan, Z.-H. Lin, Y. Xiao, K. Sun, Y.-T. Wang, J.-S. Tang, J.-S. Xu, C.-F. Li, and G.-C. Guo, \href{https://doi.org/10.1364/OPTICA.6.000067}{Optica \textbf{6}, 67-71 (2019)}.

\bibitem{ero1} B. Yurke, S. L. McCall, and J. R. Klauder, \href{https://doi.org/10.1103/PhysRevA.33.4033}{Phys. Rev. A \textbf{33}, 4033 (1986)}.

\bibitem{ero2} S. F. Huelga, C. Macchiavello, T. Pellizzari, A. K. Ekert, M. B. Plenio, and J. I. Cirac, \href{https://doi.org/10.1103/PhysRevLett.79.3865}{Phys. Rev. Lett. \textbf{79}, 3865 (1997)}.

\bibitem{ur1} A. K. Pati, U. Singh, and U. Sinha, \href{https://doi.org/10.1103/PhysRevA.92.052120}{Phys. Rev. A \textbf{92}, 052120 (2015)}.

\bibitem{ur2} M. J. W. Hall, A. K. Pati, and J. Wu, \href{https://doi.org/10.1103/PhysRevA.93.052118}{Phys. Rev. A \textbf{93}, 052118, (2016)}.

\bibitem{ur3} D. Mondal, S. Bagchi, and A. K. Pati, \href{https://doi.org/10.1103/PhysRevA.95.052117}{Phys. Rev. A \textbf{95}, 052117 , (2017)}.

\bibitem{rev35} B. Yu, N. Jing, and X. Li-Jost, \href{https://doi.org/10.1103/PhysRevA.100.022116}{Phys. Rev. A \textbf{100}, 022116 (2019)}.

\bibitem{zxev35} X. Zhao, C. Zhang, \href{https://doi.org/10.3389/fphy.2022.862868}{Front. Phys. \textbf{10}, 862868. (2022)}.

\bibitem{rev36} H. F. Hofmann, \href{https://doi.org/10.1103/PhysRevA.79.033822}{Phys. Rev. A \textbf{79}, 033822 (2009)}.

\bibitem{rev37} M. Huang, R.-K. Lee, L. Zhang, S.-M. Fei, and J. Wu, \href{https://doi.org/10.1103/PhysRevLett.123.080404}{Phys. Rev. Lett. \textbf{123}, 080404 (2019)}.
\end{thebibliography}
\end{document}